\documentclass[fleqn,usenatbib]{mnras}

\usepackage{newtxtext,newtxmath}
\usepackage[T1]{fontenc}
\usepackage{ae,aecompl}
\usepackage{hyperref}

\usepackage[usenames]{xcolor}
\usepackage{color, soul}

\usepackage{graphicx}	% Including figure files
\usepackage{amsmath}	% Advanced maths commands
\usepackage{times}
\usepackage{pifont}
\title[Anatomy of Infall]{Anatomy of a Fall: Stationary and super-Keplerian spiral arms generated by accretion streamers in protostellar discs}

\author[J. Calcino et al.]{
Josh Calcino$^{1}$\thanks{Contact e-mail: \href{mailto:jcalcino@tsinghua.edu.cn}{jcalcino@tsinghua.edu.cn}},
Daniel J. Price$^{2}$,
Thomas Hilder$^{2}$, 
Valentin Christiaens$^{3,4}$, 
\newauthor
Jessica Speedie$^{5}$,
Chris W. Ormel$^{1}$
\\
$^{1}$Department of Astronomy, Tsinghua University, 30 Shuangqing Rd, 100084 Beĳing, China\\
$^{2}$School of Physics and Astronomy, Monash University, Vic 3800, Australia\\
$^{3}$Institute of Astronomy, KU Leuven, Celestijnenlaan 200D, Leuven, Belgium \\
$^{4}$Space sciences, Technologies \& Astrophysics Research (STAR) Institute, Universit\'e de Li\`ege, All\'ee du Six Ao\^ut 19c, B-4000 Sart Tilman, Belgium\\ 
$^{5}$Department of Physics \& Astronomy, University of Victoria, Victoria, BC, V8P 5C2, Canada
}
\date{Accepted XXX. Received YYY; in original form ZZZ}

\pubyear{2024}

\begin{document}
\label{firstpage}
\pagerange{\pageref{firstpage}--\pageref{lastpage}}
\maketitle

\begin{abstract}
Late-stage infall onto evolved protoplanetary discs is an important source of material and angular momentum replenishment, and disc substructures. In this paper we used 3D smoothed particle hydrodynamics simulations to model streamer-disc interactions for a prograde streamer. The initially parabolic streamer interacts with the disc material to excite disc eccentricity, which can last on the order of $10^5$ years. We found that the spiral arms the streamer excited in the disc can have a variety of pattern speeds, ranging from stationary to super-Keplerian. Spiral arms with various pattern speeds can exist simultaneously, providing a way to diagnose them in observations. Streamer induced spirals appear similar to those generated by a massive outer companion, where the pitch angle of the spiral increases towards the source of the perturbation. Additionally, the spiral arms can show large and sudden pitch angle changes. Streamer induced spirals are long-lived, lasting approximately $3-4\times$ longer than the initial streamer infall timescale ($\sim$$10^4$ years). After the initial interaction with the disc, a long lasting low $m$ azimuthal mode persists in the disc.
\end{abstract}

\begin{keywords}
protoplanetary discs ---
circumstellar matter ---
methods: numerical ---
hydrodynamics
\end{keywords}

%%%%%%%%%%%%%%%%%%%%%%%%%%%%%%%%%%%%%%%%%%%%%%%%%%

%%%%%%%%%%%%%%%%% BODY OF PAPER %%%%%%%%%%%%%%%%%%

\section{Introduction}

Star formation is messy \citep{mccrea1960,bonnell2011}. Most of the material of a molecular cloud is not used to form stars and discs but rather disperses back into the interstellar medium (ISM) during the $\sim 5\times 10^{7}$ lifetime of the molecular cloud. During the transition from a cold molecular cloud to a loosely bound open cluster, the newly formed stars and their surrounding discs can encounter material left over from the star formation process \citep[e.g. see][]{bate2012, bate2018}. Given a typical lifetime of a few $\sim$$10^6$ years \citep{Haisch2001, mamajek2009}, most protoplanetary discs will encounter left-over material from the molecular cloud in which they were formed during their lifetimes \citep{padoan2005,throop2008,klessen2010,winter2024,Padoan2024}. This material can change the angular momentum evolution of the disc \citep{thies2011, wijnen2017, kuffmeier2024, kuffmeier2024b, pelkonen2024}.

Recent observations in both scattered light \citep[][]{benisty2023,gupta2023, garufi2024} and molecular line emission \citep[e.g.][]{huang2020, huang2021, Valdivia-Mena2024} have revealed protoplanetary discs interacting with background material. Of particular interest are discs around pre-main sequence stars thought to be quite evolved, such as AB Aur \citep[estimated age $\sim 4$ Myr;][]{garufi2024}, with strong evidence of ongoing infall. GM Aur \citep{huang2021}, RU Lup \citep{huang2020}, DR Tau \citep{Mesa2022}, DO Tau \citep{huang2022}, and HL Tau \citep{garufi2022,gupta2024} are other examples of discs with spiral and arc structures associated with infalling material.
% Other examples of discs with signs of infall include, such as

The effect of infall on the observed structure of a protoplanetary disc is not well understood, despite being argued as important since the earliest theoretical studies of disc evolution \citep[e.g.][]{ulrich1976,cameron1978,lin1990}. Recently, \cite{dullemond2019} demonstrated the formation of discs through cloudlet capture and suggested this may be the origin for FU Orionis events and transition discs. \cite{Kuffmeier2020} studied the substructures present in the second generation disc in more detail, showing spiral arms are produced. Infall misaligned with respect to a present disc can also produce a misaligned and eccentric outer disc \citep{wijnen2017, kuffmeier2021, pelkonen2024}, reminiscent of what is observed around SU Aur \citep{ginksi2021}. These previous modelling works assumed a cloudlet capture model, where the initial shape of the infalling material is a large (several thousand au in radius) sphere which is slightly tidally stretched as it reaches the star and/or disc. However observations support far more elongated structures, ``streamers'', as a more suitable initial condition \citep{alves2020, pineda2020, Cacciapuoti2024, gupta2024, hales2024}.

Simulations presented in \cite{hanawa2024} recreate an elongated streamer hitting a disc. With their model, they reproduce the kinematics of the kilo-au environment around DG Tau \citep{garufi2022}. \citet{hanawa2024} do not follow the interaction of the streamer with the disc for a long duration in their simulation, and do not specifically focus on disc substructures. With substructures being routinely observed in protoplanetary discs in mm thermal emission \citep{bae2023}, molecular line emission \citep{pinte2023}, and scattered light images \citep{benisty2023}, it is crucial to understand how late-stage infall can influence the disc. Specifically, we ought to understand what substructures can be generated by infall, and if there is any way to differentiate them from substructures generated by other mechanisms, such as planet-disc interactions.

In his paper, we provide a detailed investigation of spiral arms arising from streamer-like infall onto a T Tauri-like protoplanetary disc. We structure the paper as follows: In Section~\ref{sec:method} we briefly present our hydrodynamical simulations and how we initialise our infall event. In Section~\ref{sec:res} we study in detail the morphology of the disc and the spirals produced by infall.
In Section \ref{sec:disc} we compare our results with those obtained from simulations of gravitational instability, and provide context for our results. 
We then summarise our findings in Section~\ref{sec:sum}.

\begin{figure*}
    \centering
    \includegraphics[width=1.0\linewidth]{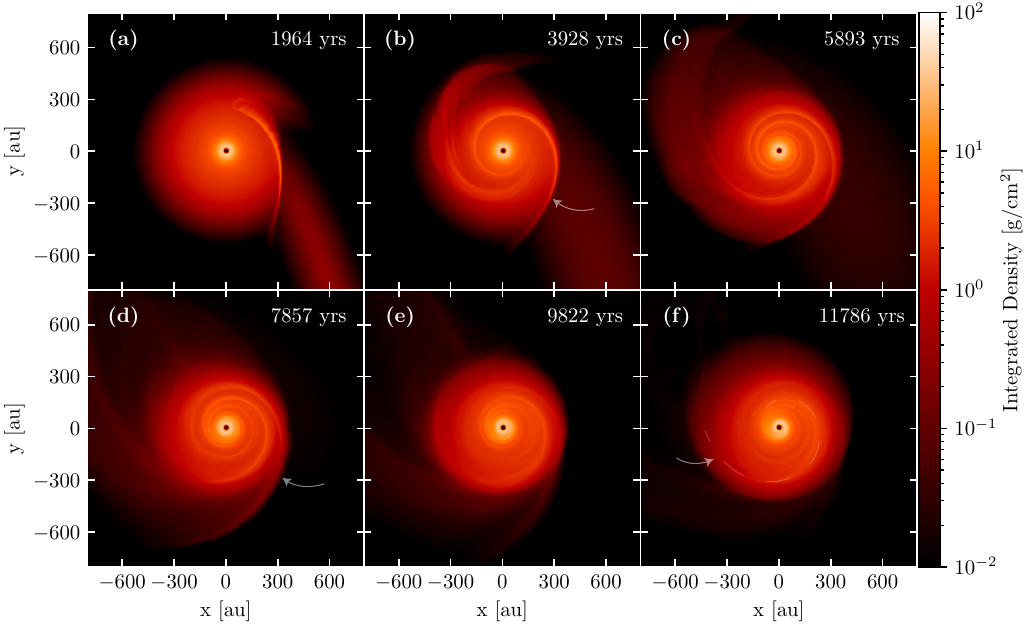}
    \caption{Simulation snapshots showing the time evolution of the streamer interacting with the disc. As the parabolic streamer material hits the disc, it launches disc material onto a higher orbit. Eventually once the streamer has completed its initial encounter with the disc, its recoiled remnants (along with the disc material it launched) re-accretes onto the disc. The initial interaction  leads to a stationary spiral in the disc (arrow in panels b and d).}
    \label{fig:time_series}
\end{figure*}

\section{Methods} \label{sec:method}

We used the smoothed particle hydrodynamics (SPH; \citealt{monaghan1992}) code {\sc phantom} \citep{phantom2018}. We simulate only one disc, to which we later add an infalling streamer. Previous studies \citep[e.g.][]{liu2016, kuffmeier2018} have shown that infall can result in the disc becoming gravitationally unstable. However in our study we neglect the disc self-gravity in order to isolate and study the structures induced by infall. We also neglect the effect of magnetic fields \citep[however see][]{unno2022}.

We initialised our gas-only simulation with $2\times 10^{6}$ SPH particles placed in the initial circumstellar disc following a surface density profile $\Sigma (R)\propto R^{-p}$ for $R_\textrm{in} < R < R_\textrm{out}$, where we set $p = 1$,  $R_\textrm{in} = 5$ au, and $R_\textrm{out} = 350$ au. We assume a locally isothermal equation of state with a sound speed profile $c_{\rm s}(r) \propto R^{-q}$ where $q=0.25$ giving $T \propto R^{-0.5}$. We set the aspect ratio of the disc $H/R_\textrm{ref} = 0.11$ at $R_\textrm{ref} = 350$ au. The total disc mass between these radii is set to $M_\textrm{disc} = 10^{-3}$ M$_\odot$. The central star is modelled as sink particle \citep{bate1995} with a $r_\textrm{acc} = 5$ au accretion radius and mass 1 M$_\odot$. Gas particles are conditionally accreted onto the sink particle if they fall between $f_\textrm{acc}r_\textrm{acc} < r < r_\textrm{acc}$ and unconditionally if $r < f_\textrm{acc} r_\textrm{acc}$, where we select $f_\textrm{acc} = 0.8$. The conditions for accretion require that accreted particles are both gravitationally bound to the sink particles and that their specific angular momentum is less than that of a Keplerian orbit at $r_\textrm{acc}$. The SPH artificial viscosity $\alpha_\textrm{AV}$ is used to produce a \citet{shakura1973} alpha viscosity $\alpha_\textrm{SS}$ following \citet{lodato2010}. We set $\alpha_\textrm{AV} = 0.1$ to give $\alpha_\textrm{SS}\approx 2.5\times 10^{-3}$. 

We first evolve the disc for $\sim$8 orbits at the outer disc radius without infall to allow it to settle any spurious structures that arise due to the disc not being in hydrostatic equilibrium. This is not quite a viscous timescale, but enough to ensure there are no more transient fluctuations in the disc. In any case, if recent studies are to be believed \citep[e.g.][]{winter2024, pelkonen2024}, discs may never truly evolve in isolation for a viscous timescale in the outer disc.

\subsection{Initialising the Infall}

We model our infalling streamer as an ellipse with an initial semi-minor axis $b$ and a semi-major axis $a$. We set $a=1000$ au and $b = 50$ au. The streamer is initialised with a mass equal to 10\% of the initial disc mass, $M_\textrm{in} = 0.001$ M$_\odot$. In this paper we only present a single model with the infall and provide a detailed study of the effects the infall has on the disc. 
The orbit of the infalling material is a parabolic orbit ($e=1$), in an almost identical fashion to the fly-by simulations by \citet[][see their Appendix~A]{cuello2019}, which has been used in previous studies \citep[e.g.][]{Borchert2022, smallwood2023}. The centre of the ellipse traces the parabolic trajectory of the initially prescribed parabolic orbit. The parabolic orbit has a periastron distance from the central star, $r_\textrm{peri}$, which we set to $r_\textrm{peri}=100$ au. We set an initial infall radius, $r_\textrm{init}$, from which the centre of the ellipse is defined. The ellipse is then distorted so that the line running along the major axis lies directly on top of the parabolic trajectory. The velocity of each particle is then set so that it is equal to the freefall velocity along the parabola, ignoring any change in velocity due to the particles being offset from the parabola due to the minor axis of the ellipse. This leads to half of the particles being bound, while the other half are unbound. 
The assumed streamer length and mass corresponds to an average accretion rate onto the disc of $\dot{M}_\textrm{in} \approx 10^{-7}$ M$_\odot$/yr. Reducing the mass further leads to the streamer not being well resolved in our SPH simulations.

\section{Results} \label{sec:res}

\subsection{Streamer-Disc Interaction}\label{sec:sd-int}

\begin{figure*}
    \centering
    \includegraphics[width=0.9\linewidth]{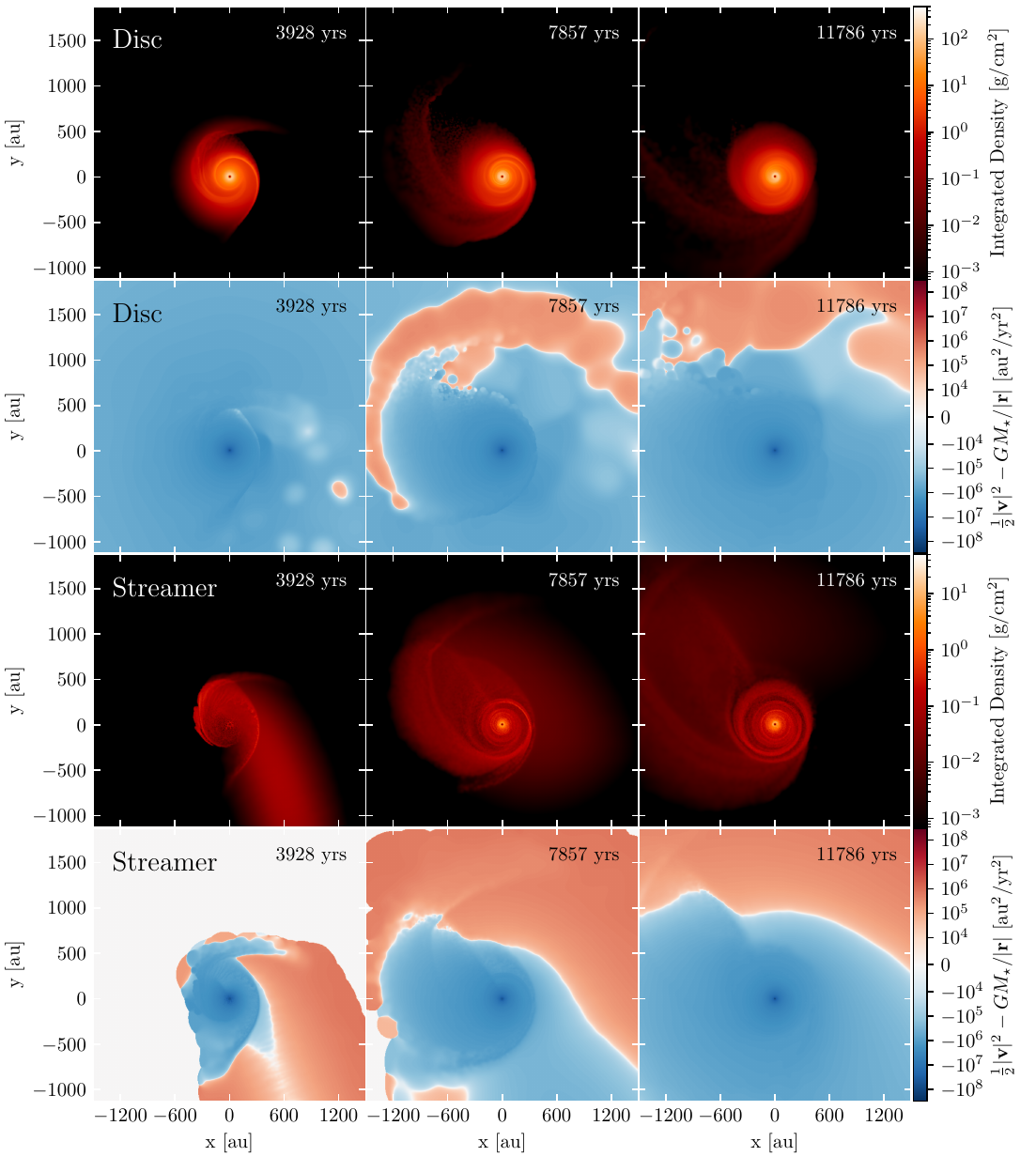}
    \caption{Simulation snapshots showing the time evolution of the streamer interacting with the disc. We have split the particles between those originally in the disc, and those originally in the streamer. The panel below the surface density plot shows the binding energy of a selection of SPH particles. For the disc, almost all particles are bound in the first snapshot. However, unbound streamer material results in part of the disc material also becoming unbound. For the streamer, most material is initially slightly unbound, but a large majority becomes bound after collision with the disc.}
    \label{fig:binding_energy}
\end{figure*}

\begin{figure}
    \centering
    \includegraphics[width=1.0\linewidth]{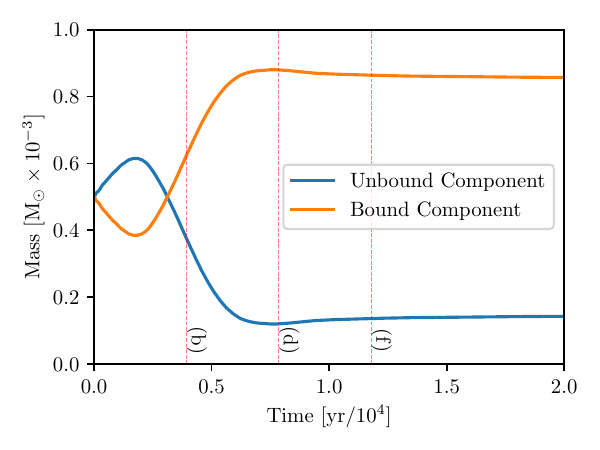}
    \caption{The total mass of bound and unbound streamer material. Initially, the streamer is perfectly split between 50\% bound and 50\% unbound. As the streamer evolves, its internal pressure is converted to kinetic energy and a larger portion becomes unbound. Most of the streamer material becomes bound when it collides with the disc. Vertical lines mark the time of the panels b, d, and f of \protect{Figure~\ref{fig:time_series}}, which are the snapshot times plotted in \protect{Figure ~\ref{fig:binding_energy}}.}
    \label{fig:bound_mass}
\end{figure}

Figure~\ref{fig:time_series} plots snapshots of the projected density for our simulation, showing the time-evolution of the streamer-disc interaction. Each panel is labelled sequentially from (a) to (f), where the later labels indicate increasingly advanced stages of the simulation. We leave a detailed analysis of the spiral structures for Section \ref{sec:spirals}, and focus here on their general behaviour. In panel (a) we see the initial stages of the streamer interacting with the disc. The initial interaction of the streamer with the disc excites a wave which propagates radially towards the inner disc in the subsequent panels (b) and (c). We mark this spiral arm with an arrow in panels (b) and (d). This initial spiral does not appear to change its position angle substantially through to panel (d). By panel (e) the streamer has essentially completed its initial interaction with the disc. However, due to the collision with bound material in the disc, much of the streamer material remains on a bound and highly eccentric orbit. 

To better illustrate the origin of material, Figure~\ref{fig:binding_energy} re-plots panels (b), (d), and (f). We specifically plot the SPH particles that are initially part of the disc or the streamer in the upper and lower half of the Figure, labelling the respective particles that are plotted as `Disc' and `Streamer', respectively. Below the integrated density projections of the disc and streamer particles, we plot the density weighted binding energy, with blue showing bound particles and red showing unbound particles.

We first focus our attention to the streamer material in Figure~\ref{fig:binding_energy}. Our streamer is initialised such that half the streamer material is initially bound, while the other half is unbound. However due to the internal pressure of the streamer, much of the particles gain kinetic energy and become unbound as the streamer approaches the disc, as seen in the streamer particles in the first column of Figure~\ref{fig:binding_energy}. As streamer material collides with the disc it loses kinetic energy and becomes bound. To illustrate the proportions of bound and unbound material, in Figure~\ref{fig:bound_mass} we plot the total mass of bound and unbound streamer particles as a function of time and include vertical lines that mark the snapshot times of panels (b), (d), and (f). We see that initially exactly 50\% of the material is unbound, and this percentage increases owing to the streamer material gaining kinetic energy which is directly converted from the internal pressure of the streamer. The collision of the streamer material with the disc at $t\sim 2\times 10^{3}$ yrs causes the portion of bound streamer material to sharply increase. By the time of snapshot (d), the portion of bound material starts to slowly decrease as material launched on near unbound orbits becomes unbound due to residual internal pressure being converted into kinetic energy.

Focusing now on the disc particles, we can see the first column of Figure~\ref{fig:binding_energy} that almost all of the disc particles are bound to the disc during the initial interaction with the streamer. Eventually a larger portion of the particles become unbound due to the incoming unbound streamer material. However most of the disc material that is strongly perturbed remains bound, though now on a highly eccentric orbit, and mixes with the streamer material. This `recoiled' material continues to re-accrete onto the disc and is responsible for some of the spiral structure that remains in panel (f) of Figure~\ref{fig:time_series}. However, by this point the disc is eccentric and another spiral-like feature arises due to this, which we discuss more in Sections~\ref{sec:spirals_speed} and \ref{sec:longterm}. Spiral structures are evident for at least $10^4$ years in our simulation, however this timescale is increased substantially due to both the re-accretion of the recoiled streamer/disc material and disc eccentricity. At the times shown in Figure~\ref{fig:time_series}, the spiral arm morphology looks similar to the spiral arms generated by a multi-Jupiter mass planet exterior to them \citep{dong2015b}.

\subsection{Infall Induced Spiral Arms}\label{sec:spirals}
\begin{figure*}
    \centering
    \includegraphics[width=1.0\linewidth]{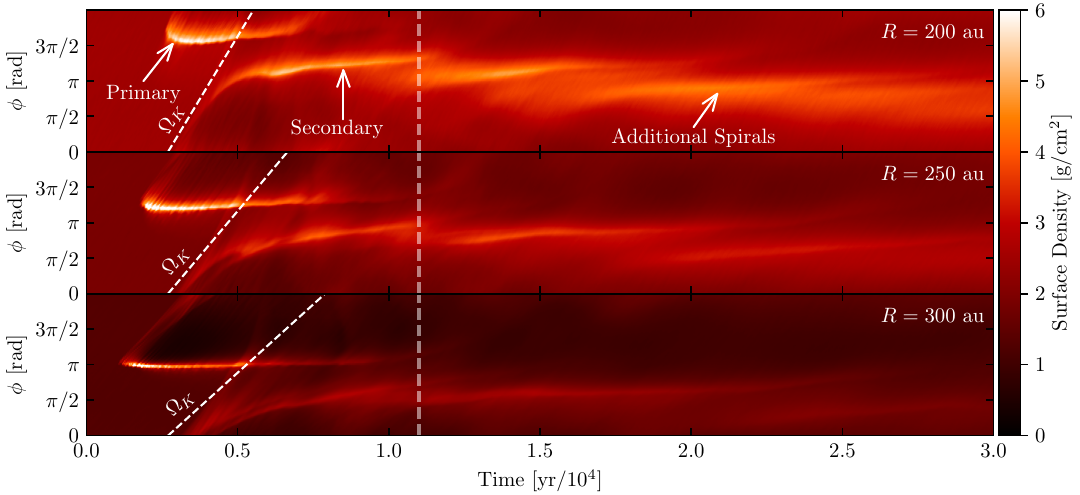}
    \caption{Three azimuthal slices at radii 200 au (top panel), 250 au (middle panel), and 300 au (bottom panel) plotted as a function of time. The vertical grey line marks the approximate end of the streamer's initial interaction with the disc. The slanted dashed line in each panel represents the orbital period assuming Keplerian rotation and negligible accretion at that particular radial distance. Also marked are the `Primary' and `Secondary' stationary spirals. The fact these are indeed stationary is evidenced by their relatively stable azimuthal location as a function of time. }
    \label{fig:standing_spiral}
\end{figure*}

\begin{figure}
    \centering
    \includegraphics[width=1.0\linewidth]{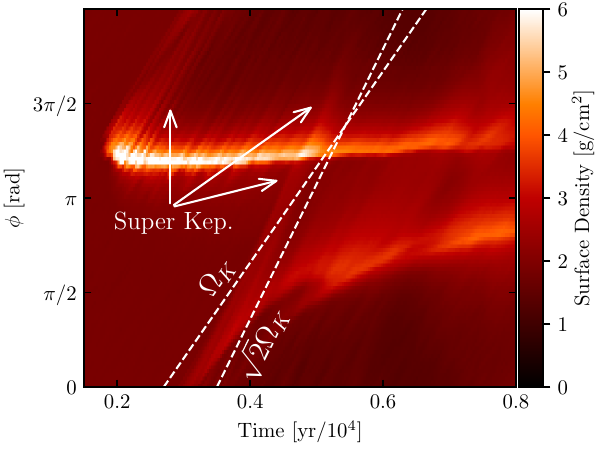}
    \caption{As in {\protect Figure~\ref{fig:standing_spiral}} but for the $R=250$ au slice and a more narrow region in time to emphasise the spiral arms moving at a super Keplerian rotation rate (marked with arrows). We plot a dashed line which has a gradient of $\Omega_K$, where its spanned $x$-axis distance represents one Keplerian orbital period. We see that the super Keplerian spirals have a gradient steeper than the Keplerian line, demonstrating their super Keplerian rotation. We also plot an additional line which shows the velocity of material falling on a parabolic orbit (which is equal to the escape velocity at that radius), assuming the motion is all in the azimuthal direction. We can see that the super Keplerian spirals move at a similar rate to the parabolic velocity.}
    \label{fig:super_kep}
\end{figure}

\begin{figure}
    \centering
    \includegraphics[width=0.9\linewidth]{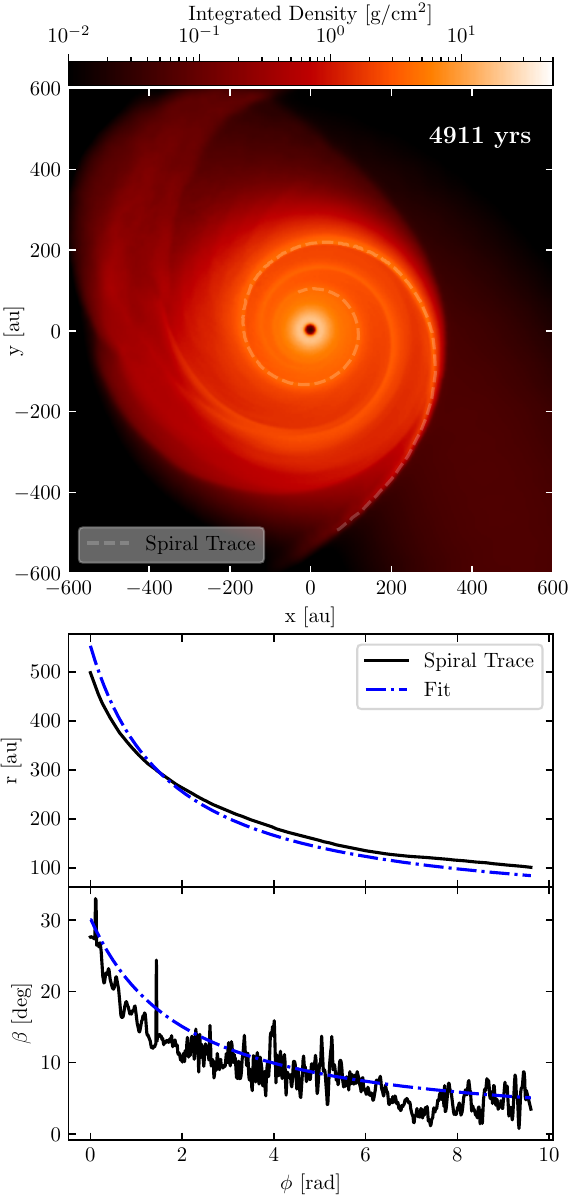}
    \caption{The properties of the stationary spiral. \emph{Top:} Surface density plot with the spiral traced by {\protect {\sc nautilus}} over-plotted. \emph{Bottom:} The properties of the spiral. The spiral plotted as a function of radial distance from the central sink against azimuthal angle $\phi$ defined from the outer tip of the spiral ($\phi = 0$ is the tip of the outer spiral trace in surface density plot) and positive in the counter clock-wise direction (middle panel, black solid line). Also plotted is the spiral pitch angle (defined in Equation~\ref{eq:beta}) as a function of $\phi$ (bottom panel, black solid line). The pitch angle of the spiral decreases as it moves away from the source of the perturbation, which is close to $\phi=0$. The form of the stationary spiral is fit well using a hyperbolic spiral (shown with a blue dot-dashed line).}
    \label{fig:spiral_prop}
\end{figure}

\begin{figure*}
    \centering
    \includegraphics[width=1.0\linewidth]{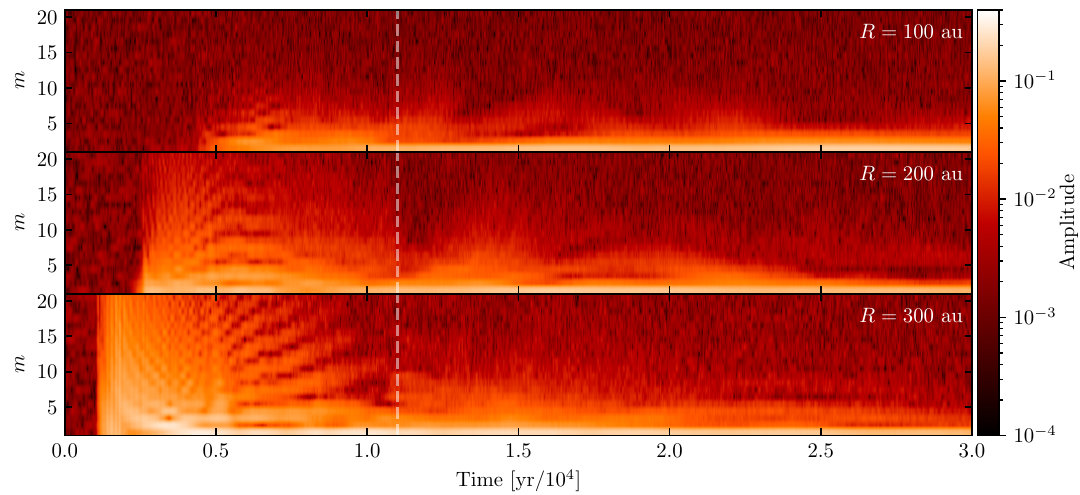}
    \caption{The amplitude of azimuthal wavenumber $m$, $A_m$, in three rings location at 100 au (top), 200 au (middle), and 300 au (bottom), plotted as a function of time. Large mode numbers are generated, particularly in the outer disc. These are damped once the infalling material has completed its initial interaction with the disc ($\sim$10$^{4}$ years, marked with grey dashed vertical line), leaving low $m$ modes to dominate.}
    \label{fig:az_modes}
\end{figure*}

\subsubsection{Stationary and Super-Keplerian Spiral Arms}\label{sec:spirals_speed}

Spiral arms are generated by the streamer as it interacts with the disc, and although their morphology looks similar to spiral arms generated by companion-disc interactions, their temporal behaviour is atypical compared with spirals generated by this mechanism. In Figure~\ref{fig:time_series} we can see that the spiral arm first formed in panel (a) does not appear to change its position angle in the disc for at least $\approx 5\times 10^{3}$ years (i.e. the time frame the streamer interacts with the disc), despite the Keplerian orbital period at 300~au being about the same order. Thus we find that streamers can induce \emph{stationary} spiral arms in the disc which propagate in the radial direction away from the interaction site between the streamer and the disc. To better illustrate these stationary spiral arms, Figure~\ref{fig:standing_spiral} shows radial slices of the disc surface density at different radial slices are a function of time. We do this for three radial separations from the central sink $R = [200, 250, 300]$ au.  A time of zero indicates the initialisation of the streamer in the simulation, and the dashed vertical grey line marks the approximate time at which the streamer has completed its initial interaction with the disc. In each row of Figure~\ref{fig:standing_spiral} a white dashed line indicates the orbital period assuming Keplerian rotation and negligible accretion at the relevant radial separation. We can see that for all radial locations sampled, the stationary spiral arms generated by the streamer is the highest amplitude perturbation in the disc surface density, and indeed appears to not move in the azimuthal direction. 

In the top panel of Figure~\ref{fig:standing_spiral} we label the `Primary' stationary spiral and `Secondary' stationary spirals, which are generated by the initial streamer material, and its recoiled remnant material, respectively. After the secondary stationary spiral, additional stationary spirals are generated, though at a lower amplitude than the first two. From Figure~\ref{fig:standing_spiral} we can see that the stationary spirals last approximately $3\times 10^{4}$ years, but they are still apparent for approximately a factor of 2 longer (not plotted). Interestingly, in Figure~\ref{fig:standing_spiral} there are also spiral structures moving at speeds exceeding the local circular Keplerian orbital speed. The increased spiral propagation speed cannot be explained by eccentricity growth in the disc, since the maximum eccentricity at the radial distances plotted in Figure \ref{fig:standing_spiral} is only $e\approx 0.15$, meaning the maximum azimuthal velocity should only be 15\% higher than in the circular case.

To more clearly demonstrate an example of this, in Figure~\ref{fig:super_kep} we plot a zoom in of the $R=250\ \textrm{au}$ slice in Figure~\ref{fig:standing_spiral}. We label a few spiral features that are moving faster than the Keplerian velocity. Super-Keplerian spirals in a disc with infall is not too surprising if we consider that the streamer itself can move with an angular velocity greater than Keplerian when it hits the disc. Material moving on a parabolic orbit has a velocity equivalent to the escape velocity at every radius, which is equal to $\sqrt{2} \Omega_K$. When plotting this slope in Figure~\ref{fig:super_kep} we can see it matches the rotation rate of the spirals in the simulation. 
We can use first appearance of the stationary wave at each radial distance to estimate its propagation speed in the radial direction. By visual inspection we see that the spiral arm propagates 50 au in $\sim 750$ years, giving a propagation speed of $\sim 320$ m/s, or approximately the sound speed.

Finally, additional stationary spiral-like structures persist in the disc long after the streamer has completed its interaction, which we label in Figure~\ref{fig:standing_spiral}. 
Some of these spirals are due to re-accreting streamer/disc material, however some spiral-like structure still remains in the disc even after removing the re-accreting material from the simulation.\footnote{We do this by removing material at a radius greater than 1000 AU at a time $\sim$2.5$\times 10^{4}$ yrs after the infall is initiated. We also tested by removing material at 750 AU and beyond at an earlier time and the resulting difference with the original test was negligible.} This demonstrates that some of these spiral arms are generated from processes within the disc itself.

\subsubsection{Pitch Angle Analysis}

We more closely analyse the structure of the stationary spiral arm in Figure~\ref{fig:spiral_prop}. We obtain points on the spiral using the code {\sc nautilus}\footnote{\url{www.github.com/TomHilder/nautilus}} and overplot them on the surface density projection in the upper half of Figure~\ref{fig:spiral_prop}. {\sc nautilus} uses a simple peak finding algorithm in the radial direction to extract points along the spiral. The radial density profile is first smoothed using a Savitsky-Golay filter, and points corresponding to local peaks are extracted with some adjustable criteria for the width and prominence of the peak. Spurious points are rejected using an estimate of the second radial derivative of the density, since it should be negative for true peaks. In the top panel of the bottom half of Figure~\ref{fig:spiral_prop} we show the radial distance of the spiral as a function of azimuthal angle, $\phi$, which is defined as being positive in a counter clock-wise direction with $\phi=0$ marking the outer tip of the spiral. The bottom panel shows the pitch angle of the spiral, defined as \citep[e.g. see][]{binney2008}
\begin{equation}\label{eq:beta}
    \beta = \left| \frac{dr}{rd\phi} \right|.
\end{equation}
The pitch angle of the spiral arm decreases as a function of $\phi$, where small $\phi$ is close to the excitation location of the spiral due to the streamer. This is qualitatively similar to planet-induced spirals \citep{bae2018}, which also increase in pitch angle close to the source of the perturbation (in this case the planet). The bottom half Figure~\ref{fig:spiral_prop} also includes our best fit to the spiral structure using a hyperbolic spiral, which is given by 
\begin{equation}
    r = \frac{a}{\phi + \phi_\textrm{shift}},
\end{equation}
where $a$ is a scaling constant and $\phi_\textrm{shift}$ azimuthally offsets $\phi$. The pitch angle of a hyperbolic spiral decreases with radius so is a suitable spiral function to use for fitting the stationary spiral arm in our simulation. 
Given this spiral is stationary, it should be easily distinguished from a planet-induced spiral with multiple scattered light observations over a sufficiently long baseline to measure the spiral arm rotation rate \cite[e.g. as in ][]{ren2018, ren2020}.

\subsubsection{Fourier Analysis}\label{sec:fourier}

We further explore and characterise the spiral morphology in the disc by conducting a Fourier analysis. We specifically focus on the amplitude of the azimuthal modes, $A_m$, which are given by 
\begin{equation}\label{eq:am}
    A_m = \frac{1}{N_{\textrm{Ann}}} \left| \sum^{N_{\textrm{Ann}}}_{i} e^{-im\phi} \right|,
\end{equation}
where $N_{\textrm{Ann}}$ is the number of particles in some azimuthal annulus, and $m$ is the azimuthal wavenumber. We compute $A_m$ as a function of time at three radial locations, 100 au, 200 au, and 300 au, and show the result in Figure~\ref{fig:az_modes}. 
At the onset of the streamer-disc interaction, many azimuthal modes are excited. As time progresses the modes are excited in the inner disc, however the higher mode numbers are increasingly damped out. The banding pattern suggests that the amplitude of the higher modes may be mostly due to harmonics and not direct excitation due to the streamer. After the streamer has completed its initial interaction with the disc at $t\sim 10^4$ years, the higher mode numbers in both the inner and outer disc are quickly damped out. At this time the smaller wavenumbers begin to dominate. They continue to dominate well after the completion of the initial streamer-disc interaction, with $m=1$ being the most dominant. The dominant $m=1$ mode is the result of the long lasting spiral structures discussed in Section~\ref{sec:spirals_speed}.
%In the surface density 

\subsection{Long Term Evolution}\label{sec:longterm}
\subsubsection{Disc Evolution}

\begin{figure*}
    \centering
    \includegraphics[width=1.0\linewidth]{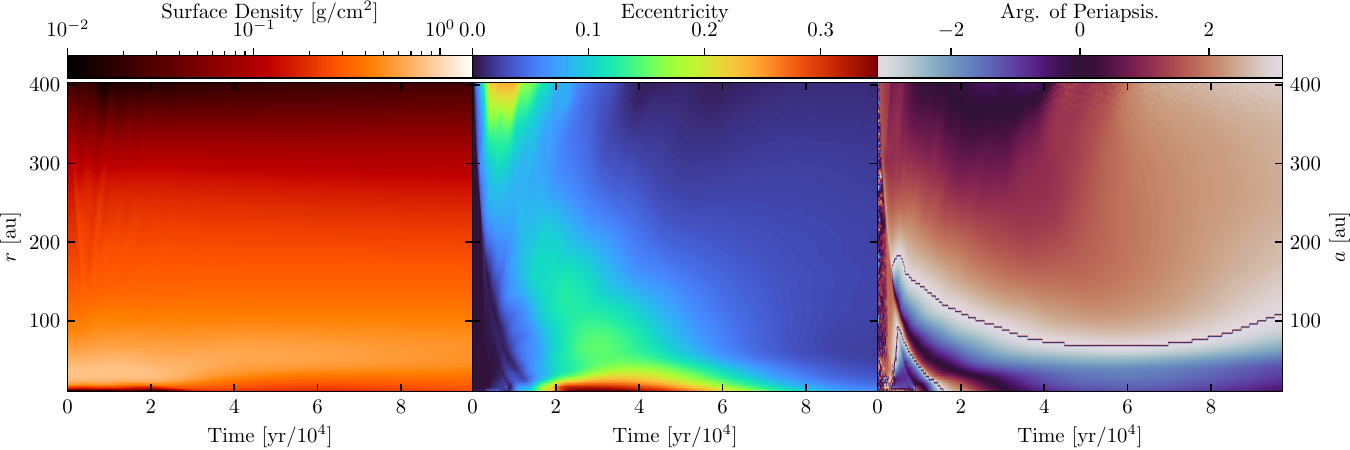}
    \caption{Azimuthal averages of surface density (left panel), disc eccentricity (middle panel), and argument of periapsis $\omega$ (right panel), as a function of time. The surface density is azimuthally averaged as a function of $r$, while the eccentricity and argument of periapsis are azimuthally averaged with respect to semi-major axis $a$. Spiral structure is evident in the azimuthal surface density for the first $\sim 3\times 10^{4}$ years, though they are evident for longer in regular surface density projections. Material close to the central sink is initially depleted, but the region eventually refills. As the streamer hits the outer disc it excites eccentricity which propagates towards the inner disc, creating a large spike in eccentricity around the sink as this region refills with material. An elevated eccentricity of $e\sim 0.05$ persists throughout the disc for at least $10^{5}$ years, with the inner regions retaining the highest eccentricity. As the disc eccentricity grows, $\omega$ begins to show a dependence on $a$ leading to the development of a twisted disc. }
    \label{fig:ecc_time}
\end{figure*}

\begin{figure}
    \centering
    \includegraphics[width=0.9\linewidth]{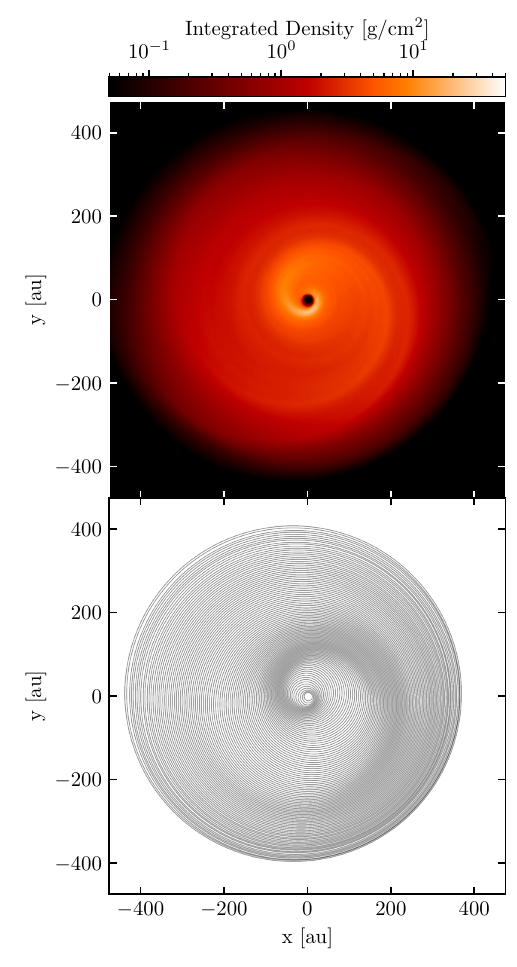}
    \caption{A snapshot of the surface density of our simulation at $t\sim 2\times10^{4}$ years after the addition of the streamer (top panel). We plot concentric ellipses obtained from the azimuthally averaged (in bins of semi-major axis $a$) eccentricity vector in the bottom panel. Due to the changes in the argument of periapsis $\omega$ in the eccentric disc, spiral-like structures appear due to a pile-up of the ellipses. This shows us that the long lasting spiral and low $m$ mode is at least in part due to the disc eccentricity.}
    \label{fig:ecc_disc}
\end{figure}

% 1. Discuss the eccentricity damping

Our initially smooth Keplerian accretion disc is strongly perturbed by the streamer material, and in this Section we further analyse the long term effects on the disc. 
In Figure~\ref{fig:ecc_time} we plot azimuthal averages in the range $ 10\ \textrm{au} \leq r \leq 400 \textrm{au}$ of the surface density in the left panel. In the middle panel, we plot the disc eccentricity azimuthally averaged over the semi-major axis $a$ over the range $ 10\ \textrm{au} \leq a \leq 400\ \textrm{au}$. We compute the disc eccentricity from the eccentricity vector and average over the semi-major axis rather than radius, since this eliminates spurious eccentricities that arise due to the gas pressure support \citep{ragusa2018}. Specifically, we follow the process specified in \cite{Teyssandier2017}. In the right panel of Figure~\ref{fig:ecc_time} we plot the argument of periapsis obtained from the eccentricity vector.

Perturbations in the azimuthally averaged surface density are evident due to the spiral structure generated, but disappear in the azimuthally averaged profile by roughly $t \sim 3 \times 10^4$ years (though are evident for longer periods in surface density plots). Eccentricity in the disc is large in two regions. Firstly, the azimuthally averaged eccentricity profile shows that that as soon as the streamer material interacts with the outer disc it becomes eccentric. This can be qualitatively understood with our discussion of the general behaviour of the streamer material in Section \ref{sec:sd-int}. The initially parabolic streamer material interacts with disc material and drives the disc eccentricity. This eccentricity then propagates from the outside-in. However we also see large eccentricity generated in material close to the central sink. It is not clear what is causing the growth in eccentricity close to the sink. 
In the azimuthally averaged surface density profile we see that material is relatively depleted close to the sink. By $\sim 3\times 10^4$ years this depleted region is filled with material, at roughly the same time the eccentricity close to the sink reaches a maximum. Therefore, the large eccentricity close to the sink may be due to this depleted region being refilled with material that is on an eccentric orbit itself. Interestingly, the eccentricity in the outer disc appears to damp faster than the inner disc. This may be due to an increased artificial viscosity owing to the lower resolution in the outer disc.

In the right panel of Figure~\ref{fig:ecc_time} we plot the argument of periapsis, $\omega$. At the beginning of the simulation, $\omega$ is uncorrelated with $a$. However once eccentricity is induced in the disc by the infall, $\omega$ begins to show a dependence on $a$, leading to a `twisted' disc \citep{ogilvie2014}. The pile-up of the apocenters of concentric eccentric rings result in apparent spiral structure which explains our dominant $m=1$ mode seen long after the infall has finished. To demonstrate this, we take a snapshot of our simulations at $t\sim 3\times10^4$ years and obtain the azimuthally averaged eccentricity vector and $\omega$ profiles. We then plot the surface density of the simulation in the top panel of Figure~\ref{fig:ecc_disc}. In the bottom panel we plot the concentric ellipses obtained from the azimuthally averaged eccentricity and $\omega$. Hence, these ellipses are simply generated from the azimuthally averaged eccentricity profile of the disc. A spiral-like feature arises due to the twisting of the eccentricity vector as a function of radius \citep[e.g. see][]{ogilvie2014, ragusa2018} producing a pile-up of eccentric material at the apoapsis of the ellipses. This pile-up of eccentric material coincides with the spiral-like feature seen in the surface density plot, demonstrating it is generated by a pile-up of the ellipses. However, there still appears to be additional, albeit low contrast, spiral arms which could be due to the residual infall material interacting with the disc. Therefore, spiral arms are generated both by the infalling material launching waves in the disc, and also by the pile-up of eccentric material in the disc. Both of these mechanisms produce spiral features that have pattern speeds substantially slower than the local Keplerian value, explaining the additional stationary spiral structures in the right side of Figure~\ref{fig:standing_spiral}.

\subsubsection{Accretion}

\begin{figure}
    \centering
    \includegraphics[width=1.0\linewidth]{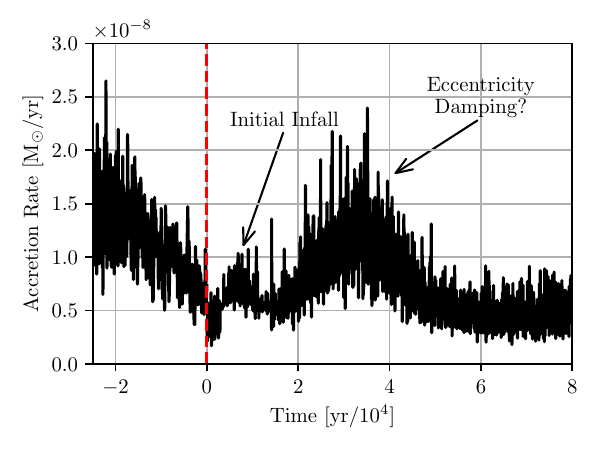}
    \caption{Accretion rate as a function of time. In this plot, a time of zero indicates the initialisation of the streamer in the simulation. We see that the accretion rate onto the sink does not increase substantially during the first $\sim 2\times 10^4$ years, despite the disc displaying obvious spiral structure due to the streamer-disc interaction. The accretion rate peaks at $3\times 10^4$ years, but only roughly a factor of 2 higher than the accretion rate seen when the streamer was initialised. The prograde and co-planar streamer in our simulation does not produce a large accretion outburst.}
    \label{fig:mdot}
\end{figure}

In Figure~\ref{fig:mdot} we plot the mass accretion rate onto the central star as a function of time. In this Figure, negative times indicate the times prior to the streamer being added to the simulation. Although the accretion rate has not reached quasi-equilibrium prior to our injection of the infalling material, we see that it is on a downwards trend. Neglecting any outside perturbation, this trend would continue inline with predictions of viscous disc evolution \citep[e.g. see][]{pringle1981}. Therefore we can attribute any increase in the accretion rate to the infalling material.

We see in Figure~\ref{fig:mdot} that the accretion rate peaks at roughly $\sim 3\times 10^4$ years, just as the eccentricity close to the sink reaches a maximum. The streamer does not substantially increase the mass accretion rate onto the star until some time after its initiation, and even then by only a factor of $\sim$2. For a parabolic orbit, the angular momentum of the streamer at its closest approach ($r_\textrm{in}$) is a factor of $\sqrt{2}$ higher than the disc material at the same radius if the streamer is co-planar with the disc. Therefore it is not surprising that the accretion rate onto the sink does not immediately increase. The peak in accretion rate at $3\times 10^4$ years is not due to the recoiled streamer material. We tested this by removing streamer material outside the disc, as described in Section \ref{sec:spirals_speed}. With the streamer material removed the accretion rate onto the sink does not substantially change, nor does the eccentricity damping timescale. The peak in accretion rate is then likely due to the eccentricity damping, rather than the recoiled streamer material re-accreting onto the disc. 

\subsection{The Kilo-au Environment and Peculiar Spirals}\label{sec:kilo-au}

\begin{figure}
    \centering
    \includegraphics[width=1.0\linewidth]{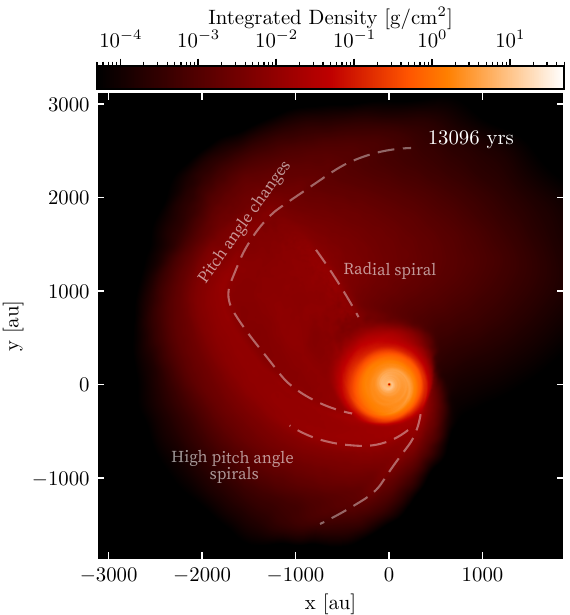}
    \caption{The kilo-au environment around the disc after the streamer has completed its initial interaction with the disc. Spiral structures with large and sudden pitch angle changes are evident in the environment surrounding the disc.}
    \label{fig:kiloau}
\end{figure}

After the streamer hits the disc, much of the streamer material and some of the disc material is sent onto a high orbit, or is unbound altogether. Figure~\ref{fig:kiloau} shows the projected surface density several thousands of au away from the central sink to display the kilo-au environment after the infall event. Of particular interest are the multiple high pitch angle spirals towards the south, and large scale spiral arm spanning several thousand au which displays a sharp pitch angle change. This high pitch angle change is apparent when the streamer first interacts with the disc (Figures \ref{fig:time_series} and \ref{fig:binding_energy}). Leading this large scale spiral arm are two additional spiral arms which have a high pitch angle. We also see a spiral arm which appears almost entirely radial. 
This general structure with abrupt pitch angle changes and multiple spirals persists for at least $\sim 5\times 10^4$ years in our simulation.

The mechanism of producing these abrupt changes in pitch angle is unclear, however we provide a qualitative description of how some of the spiral arms are produced in our simulation. By referring to Figure~\ref{fig:binding_energy}, we see that what eventually becomes the `radial spiral' is only apparent in the streamer particles. It is produced due to some of the streamer material falling on the opposite side of the sink particle compared with the bulk streamer, causing to it have a retrograde orbit. This material then collides with material on a prograde orbit, cancelling out their orbital motion. The radial spiral appears to move away from the sink with time. Additional streamer material that ends up on a retrograde orbit collides with diffuse prograde disc material. This appears to be the origin of the leading high pitch angle spiral. Inspecting subplot (d) of Figure~\ref{fig:time_series} we see that this leading high pitch angle spiral appears to be an extension of the stationary spiral described in Section \ref{sec:spirals}. The trailing high pitch angle spirals results from disc and streamer material that remains bound to the star, but is on a high and eccentric orbit.

We may then begin to speculate on the cause of the large pitch angle change. Once again inspecting Figure~\ref{fig:binding_energy}, we observe that the abrupt pitch angle change is formed close to the intersection with the radial spiral. Some infalling and disc material on the prograde side of the radial spiral (right side in our projection) is unbound. Simultaneously, streamer and disc material is also bound and orbiting in the prograde direction, albeit on a high orbit with large eccentricity. The intersection between these two zones is exactly where the large pitch angle change occurs, as can be seen in the binding energy plots of Figure~\ref{fig:binding_energy}. Therefore the large pitch angle changes are likely the result of streamer material changing the binding energy (and angular momentum) of material along the location where it hits the disc, as argued by \citet{cameron1978}.

\section{Discussion}\label{sec:disc}

\subsection{Streamer-Induced Substructures}

Previous studies have shown infall can produce a variety of disc substructures. The work by \cite{bae2015} using the isothermal and spherically symmetric Ulrich, Cassen, and Moosman \citep[UCM][]{ulrich1976,cassen1981} model found spiral arms and vortices are excited. In these 2D simulations, the infall was added as a source term in the continuity equation between $0 \leq R \leq R_c$, where $R_c$ is the centrifugal radius. A vortex was found to be generated near $R_c$ due to a sharp change in vortensity. Later simulations by \cite{Kuznetsova2022} used a modified version of the UCM model to include a minimum infall radius $R_\textrm{in} \leq R \leq R_c$ and found that vortices could also be produced at the inner transition zone of the infall. Both works demonstrated that along with spiral arms, dust rings and horseshoes can be produced. They also showed that infall and the spirals it produces result in angular momentum transport that can be parametrised using the $\alpha$-prescription \citep[see also][]{lesur2015,Hennebelle2016}. However both works assumed axisymmetric infall, which may not be a valid assumption even for the early stages of disc formation \citep{commercon2024}. \cite{Hennebelle2016} also performed 2D simulations with infall, but injected material into the simulation near the outer boundary (testing both symmetric and asymmetric injections) rather than modifying the source term in an axisymmetric fashion as in \cite{bae2015}. Vortices were not produced in their work likely due to their assumption of purely radial infall which excited both the Rossby-Wave instability and the Rayleigh centrifugal instability.

The residual viscosity caused by our adoption of a constant $\alpha_{\rm av}$ in the disc may hinder the formation of vortices in our simulation. However, given that even inviscid grid-based simulations with asymmetric infall do not tend to produce vortices, we may not expect them to form in our simulation even with a lower numerical viscosity.

Other works in the literature have found non-Keplerian motion in the recipient disc due to infall. The work by \cite{Hennebelle2017} studied non-axisymmetric infall using 3D simulations. Similar to our work, they also found that spiral arms with little proper motion are produced by asymmetric infall. The study by \citet{unno2022} also found magnetically driven sub- and super-Keplerian motion in the disc induced by a magnetised infall event. Although they did not measure spiral arm pattern speeds in their study, these non-Keplerian motions likely also result in spiral arms with non-Keplerian pattern speeds.

% \textbf{Spiral arms induced by infall were also reported to have little proper motion by \citep{Hennebelle2017}

\subsection{Streamer-Induced vs GI Induced Spirals}\label{sec:inf_vs_gi}

The spiral structure generated by infalling material is reminiscent of those produced by the gravitational instability (GI), particularly when the streamer and disc are interacting. Our Fourier analysis in Section \ref{sec:fourier} is similar to that conducted on GI discs by \cite{cossins2009}, allowing us to draw some comparisons. They find that for GI-driven spiral structure, $m\sim 1/q_R$ where $q_R$ is the mass ratio between the disc and the star, meaning that the average azimuthal mode decreases with an increasing disc to star mass ratio. At times the most dominant azimuthal mode peaks around the same value in our simulation, however smaller mode numbers dominate for most times. This may change if we change our initial conditions, for example, by including multiple streamers, or adding turbulence to the streamer, which may result in the streamer being more fragmented and landing less coherently on the disc. Our simplified setup here at least shows us distinguishing between spiral substructures driven purely by streamers or purely by GI on the spiral arm morphology alone will be difficult. This is complicated by the fact that we do not observe the disc surface density directly and rely on molecular tracers (which are observed with finite spatial/spectral resolution and noise) to reconstruct it. We also stress that infall and the gravitational instability are not mutually exclusive phenomena, with simulations showing that infall often leads to gravitational instability \citep{liu2016, kuffmeier2018}. This concept also has observational support in the case of infall fuelling the GI-induced spirals in the disc around AB~Aur \citep[][Speedie et al. submitted]{fukagawa2004,dullemond2019,speedie2024}. Our results are relevant for cases when the infalling material has insufficient mass to excite GI in the recipient disc. In this case we need methods of distinguishing between the substructures exclusively driven by infall with those driven by gravitational instability.
% This is made worse by the addition of finite spatial/spectral resolution and noise. 

One way of distinguishing between streamer-induced and GI-induced spiral arms is to measure their rotation rates \citep{ren2018, ren2020}. Our simulations show that streamer-induced spiral arms exhibit rotation rates ranging from zero to super-Keplerian, and these can be seen simultaneously. For a GI disc we expect the spirals to rotate at a slightly super-Keplerian speed owing to the self-gravity component from the disc, with only small differences between the rotation rate of individual spirals. 
Therefore, if observations reveal abundant spiral structures and the spirals appear to move with varying rotation rates, this strongly favours streamer-disc interactions, as opposed to GI. 

Another way of distinguishing may be through analysing the disc kinematics. As shown by \cite{hall2020}, GI discs produce characteristic wiggles in the channel maps of spectrally resolved emission lines \citep[see also][]{Longarini2021}. The review by \cite{pinte2023} showed that perturbations induced in the disc by multiple mechanisms can all leave unique signatures in the disc kinematics. Although we have not produced radiative transfer simulations and mock observations in this work, we can reasonably expect infall induced spirals to also have a kinematic signature. We leave the detailed study of infall-induced kinematic signatures to a forthcoming work.

% Detailed study of the kinematic signatures induced by GI indicate that the radial velocity component should be radially convergent \citep{hall2020, Longarini2021}. That is, the radial velocity residuals show radial motion that moves towards the centre of the spiral arm. This is in contrast with planet-induced spirals, where the radial velocity is divergent from the centre of the spiral arm \citep{goodman2001, rafikov2002, bollati2021, calcino2021}. This signature was recently identified in AB Aur by \cite{speedie2024}, suggesting a GI origin for the spirals. However AB Aur is also suspected to be interacting with the environment through infall \citep{dullemond2019, Kuffmeier2020}, albeit in a more complex way to how infall is modelled in this paper (Speedie et al. submitted, see also Section \ref{sec:lim}).
% It is not clear whether spirals induced by streamers will produce radially divergent or convergent radial velocities, or even a combination thereof. The kinematic signatures of streamer-disc interactions will be the focus of a forthcoming work.

\subsection{Observational Consequences}

Given we have not performed radiative transfer and produced mock observations in this paper, our comparison with observations must remain somewhat qualitative. The recent study by \cite{Krieger2024} show that the spiral arm structures induced by infall are observable with current facilities. As we showed in Section \ref{sec:spirals_speed}, the rotation rate of spiral arms induced by infall can range from no motion at all, to super-Keplerian, with a mix of propagation speeds in-between. Spiral arms with varying rotation rates can also co-exist. 

There are a handful of systems where the spiral arm rotation rate has been measured \citep{Juillard2022, xie2023, ren2024}, such as MWC~758 \citep{ren2018, ren2020}, HD~135344B \citep{xie2021}, and HD~100543 \citep{xie2023}, with the latter having a clear connection with an observed stellar mass companion.
For MWC~758, several claimed detections have been made for planetary mass companions \citep[e.g.][]{reggiani2018, wagner2019, wagner2023} that may be able to reproduce the disc substructures \citep[e.g. see][]{dong2015b, calcino2019, baruteau2019}. However both claimed companions are in locations that are in tension with the inferred orbit location of the perturbing body based on the analysis of spiral motion \citep{ren2020}. Although the rather slow spiral rotation rate could be suggestive of infall induced spirals, their morphology and a relatively constant pitch angle along the spiral suggests they are not being actively generated by infall in the outer disc. There is also, to our knowledge, no evidence of nebulosity around the disc or high extinction of the central star, both of which would indicate ambient material. Furthermore, the spirals appear to be rotating with similar rotation rates, whereas we would expect different rotation rates. This appears to be the case in HD~135344B \citep{xie2021}, but again there is little evidence to suggest ongoing infall in this system.

Another possibility is that these systems, or many other class II and transition discs, have experienced infall within the last $\sim$10$^5$ years. Given our prediction of eccentricity pumping in the disc, which takes a long time to damp, we may expect to still see evidence of past infall if the disc is eccentric. This appears to be the case in MWC~758 \citep{dong2018, kuo2022}, however these constraints have only been obtained close to the central cavity, where the eccentricity could be companion driven \citep[e.g.][]{ataiee13a, Teyssandier2016, Teyssandier2017, hirsh2020, calcino2023, calcino2024}. 

However, our work here makes a rather interesting prediction for discs that have experienced infall within the last $\sim$10$^5$ years. 
We should expect the outer disc to display some residual spiral-like structures owing to the disc eccentricity, with a strong $m=1$ mode due to the eccentricity vector pointing in different directions as a function of radius. 
This feature can still be present even if the deprojected disc morphology shows the outer disc is circular. Given that gas particles in an eccentric disc experience vertical motions along their orbit \citep[e.g. see][]{ragusa2024}, this $m=1$ mode could be visible in the velocity residuals of our simulation. Such a signal is seen in the disc around J1604 by \citet[][their Figure 4]{stadler2023}. Given that the disc eccentricity could last an order of magnitude or so longer than the infall event, any residual and unbound infall material may have time to disperse back into the ISM without leaving such an obvious signature. We also note that \cite{commercon2024} recently reported that discs are born eccentric due to anisotropic accretion. Late-stage infall is also anisotropic \citep{dullemond2019, Kuffmeier2020, kuffmeier2023}, and hence may be a sporadic source of disc eccentricity throughout the disc lifetime.

For other systems, such as HD~142527 and IRS~48, the disc eccentricity is larger \citep[e.g.][]{Hunziker2021, yang2023} and ALMA CO line observations display foreground contamination \citep[e.g.][]{vandermarel2015b, garg2021}. For HD~142527, the central cavity was once thought to be carved by the central binary \citep{biller2012, price2018}, however more recent constraints on the companion now suggest the binary is too compact to produce such a large cavity \citep{nowak2024}. Thus far IRS~48 has no confirmed companion, although modelling studies suggest either a planetary \citep[e.g.][]{zhu2014} or stellar \citep{ragusa2017, calcino2019} mass companion inside the cavity. It is also plausible that infall is playing a role in producing or at least affecting these transition discs \citep{dullemond2019}. In the paradigm of late-stage infall through a Bondi-Hoyle Lyttleton type process \citep{bondi1952, edgar2004-review}, higher mass stars are expected to accrete more material owing to their larger Hill sphere \citep{padoan2005, throop2008, klessen2010, smith2011, padoan2020, pelkonen2021, winter2024}, with simulations by \cite{kuffmeier2023} suggesting that stars with a mass greater than 1M$_\odot$ accrete more than 50\% of their final mass 500 kyr after formation. High mass stars are also much more likely to be binaries than low mass stars \citep[e.g. see][and references therein]{elsender2023, chen2024}. Then perhaps it is expected that the more massive stars and binary systems are the ones more likely to experience late-stage infall, and hence are more likely to display the morphological and kinematical signatures we expect from infall. For example, HD~142527 displays high pitch angle spiral arms in the outer regions of the disc \citep{fukagawa2006, Hunziker2021}, as does the circumbinary disc around HD~34700A \citep{monnier2019, Uyama2020b, chenmh2024, columba2024}. Other discs proposed to be circumbinary, such as AB~Aur \citep{poblete2020, calcino2024} and HD~100546 \citep{norfolk2022} also have rich spiral structure in their outer discs due to inferred interaction with infalling material \citep{grady1999, grady2001, fukagawa2004, ardila2007, dullemond2019}. 
 
Spiral arms with abruptly changing pitch angles are seen around several systems with infall, including DO Tau \citep{huang2022}, DR Tau \citep{Mesa2022}, and AB Aur \citep[][Speedie et al. submitted]{boccaletti2020}. Indeed, Figure 4 of \citet{Mesa2022} shows spiral structure that looks eerily similar to our Figure~\ref{fig:kiloau}, although the spatial scales in their observations are smaller than in our simulation. However since our simulations are scale-free, we could simply reduce the length scale in our simulation to produce a closer match. 

By changing the initial conditions of our simulation, by for example including multiple smaller streamers, we should be able to produce more abundant spiral structure in the disc \citep[see also][]{dullemond2019, Kuffmeier2020}. This could help to reproduce the complex spiral structure seen in discs which appear to be undergoing infall, such as AB~Aur \citep{boccaletti2020}, WW Cha, and DoAr 21 \citep{garufi2020}. %In this case, it appears that the disc is fed from multiple disc scale streamers \citep[][Speedie et al. submitted]{tang2012-abaur-envelope}.

\subsection{Limitations}\label{sec:lim}

There are limitations intrinsic to the method we use to simulate our infall, along with the prescription and physics of the infall we include. SPH is a meshless method, and is well suited for problems with multiple scales and complex geometries. However higher numerical viscosity may cause transient features induced by the infall to be damped more quickly. 
Our streamer prescription is certainly an oversimplification of reality. 
However, this is likely true of other works in the literature which assume spherical cloudlets which then interact with or produce a disc \citep[e.g.][]{dullemond2019, Kuffmeier2020, hanawa2024}. 
Our assumption of a locally isothermal equation of state may not be too unrealistic since we are focusing on Class II-like discs \citep{law2021}. Additionally we have neglected the effects of magnetic fields \citep[but see][]{unno2022}.
We must not forget that many protoplanetary discs, even in Class II, still appear spatially co-located with their natal and turbulent GMC environment \citep[e.g. see Figure 1 of][]{garufi2024}. Therefore the material infalling onto them is likely a product of that turbulent environment, and may even carry turbulence with it, leading to a cascade of turbulence from the GMC scale down to the disc scale \citep[e.g.][]{klessen2010}. The most realistic way of studying late-stage infall would be to evolve a GMC simulation until Class II-like discs are present and then follow their evolution \citep{kuffmeier2024}. Simulations currently either lack the resolution to resolve disc substructures \citep[e.g.][where the smallest gridsize is 25 au]{kuffmeier2023}, or end the simulation when most systems are still in the Class 0/I phase \citep[e.g.][]{bate2012, kuffmeier2017}. However if our primary focus is to reproduce the observed features in a particular source, simulating the evolution of an entire star forming region is inefficient, particularly if we want to include additional physics (e.g. dust grain growth and fragmentation, radiation). The inclusion of a turbulent background that produces infalling filamentary structures may be a satisfactory compromise.

\section{Summary}\label{sec:sum}

Streamer material infalling onto a protoplanetary disc can induce a variety of spiral structures with properties both similar and dissimilar to those produced by other mechanisms. It can also produce observable changes to the long-term disc evolution.
Our main findings are the following:
\begin{enumerate}
    \item Stationary spirals, which have little to no orbital motion, are generated in conjunction with spirals orbiting with a super Keplerian rotation rate. These spirals can exist simultaneously.
    \item The pitch angle of the spiral arm increases towards the source of the perturbation, similar to spiral arms generated by a planet exterior to the spiral arms \citep[e.g. see][]{dong2015b}.
    \item The streamer induces eccentricity in the disc which is long lived compared with the infall timescale in our simulations ($\sim$10$^5$ years). 
    \item Disc eccentricity pumping leads to a dominant $m=1$ mode which manifests as a spiral-like feature in the disc surface density.
    \item Spiral arms with sudden pitch angle changes are generated, and may be present either in the disc or the surrounding environment. Such sudden pitch angle changes are seen around several discs suspected of encountering late stage infall, such as DO Tau \citep{huang2022} and DR Tau \citep{Mesa2022}.
\end{enumerate}

\section*{Acknowledgements}
We thank the referee for their constructive comments which improved the quality of our manuscript. 
This work is supported by the National Natural Science Foundation of China under grant No. 12233004 and 12250610189.
Je.S. is supported by the Natural Sciences and Engineering Research Council of Canada (NSERC).
DJP is grateful for Australian Research Council funding via DP180104235, DP220103767 and DP240103290, and thanks Hans Zinnecker, Andrew Winter and Myriam Benisty for useful discussions.
VC thanks the Belgian F.R.S.-FNRS, and the Belgian Federal Science Policy Office (BELSPO) for the provision of financial support in the framework of the PRODEX Programme of the European Space Agency (ESA) under contract number 4000142531.

\section*{Data Availability Statement}
{\sc phantom} is publicly available at \url{https://github.com/danieljprice/phantom}. Scripts and data necessary to reproduce all the figures in this paper are available at \url{10.5281/zenodo.14823309}.

%%%%%%%%%%%%%%%%%%%%%%%%%%%%%%%%%%%%%%%%%%%%%%%%%%



\bibliographystyle{mnras}
\bibliography{paper} % if your bibtex file is called example.bib

\begin{thebibliography}{}
\makeatletter
\relax
\def\mn@urlcharsother{\let\do\@makeother \do\$\do\&\do\#\do\^\do\_\do\%\do\~}
\def\mn@doi{\begingroup\mn@urlcharsother \@ifnextchar [ {\mn@doi@} {\mn@doi@[]}}
\def\mn@doi@[#1]#2{\def\@tempa{#1}\ifx\@tempa\@empty \href {http://dx.doi.org/#2} {doi:#2}\else \href {http://dx.doi.org/#2} {#1}\fi \endgroup}
\def\mn@eprint#1#2{\mn@eprint@#1:#2::\@nil}
\def\mn@eprint@arXiv#1{\href {http://arxiv.org/abs/#1} {{\tt arXiv:#1}}}
\def\mn@eprint@dblp#1{\href {http://dblp.uni-trier.de/rec/bibtex/#1.xml} {dblp:#1}}
\def\mn@eprint@#1:#2:#3:#4\@nil{\def\@tempa {#1}\def\@tempb {#2}\def\@tempc {#3}\ifx \@tempc \@empty \let \@tempc \@tempb \let \@tempb \@tempa \fi \ifx \@tempb \@empty \def\@tempb {arXiv}\fi \@ifundefined {mn@eprint@\@tempb}{\@tempb:\@tempc}{\expandafter \expandafter \csname mn@eprint@\@tempb\endcsname \expandafter{\@tempc}}}

\bibitem[\protect\citeauthoryear{{Alves}, {Cleeves}, {Girart}, {Zhu}, {Franco}, {Zurlo}  \& {Caselli}}{{Alves} et~al.}{2020}]{alves2020}
{Alves} F.~O.,  {Cleeves} L.~I.,  {Girart} J.~M.,  {Zhu} Z.,  {Franco} G. A.~P.,  {Zurlo} A.,   {Caselli} P.,  2020, \mn@doi [\apjl] {10.3847/2041-8213/abc550}, \href {https://ui.adsabs.harvard.edu/abs/2020ApJ...904L...6A} {904, L6}

\bibitem[\protect\citeauthoryear{{Ardila}, {Golimowski}, {Krist}, {Clampin}, {Ford}  \& {Illingworth}}{{Ardila} et~al.}{2007}]{ardila2007}
{Ardila} D.~R.,  {Golimowski} D.~A.,  {Krist} J.~E.,  {Clampin} M.,  {Ford} H.~C.,   {Illingworth} G.~D.,  2007, \mn@doi [\apj] {10.1086/519296}, \href {https://ui.adsabs.harvard.edu/abs/2007ApJ...665..512A} {665, 512}

\bibitem[\protect\citeauthoryear{{Ataiee}, {Pinilla}, {Zsom}, {Dullemond}, {Dominik}  \& {Ghanbari}}{{Ataiee} et~al.}{2013}]{ataiee13a}
{Ataiee} S.,  {Pinilla} P.,  {Zsom} A.,  {Dullemond} C.~P.,  {Dominik} C.,   {Ghanbari} J.,  2013, \mn@doi [\aap] {10.1051/0004-6361/201321125}, \href {http://adsabs.harvard.edu/abs/2013A%26A...553L...3A} {553, L3}

\bibitem[\protect\citeauthoryear{{Bae} \& {Zhu}}{{Bae} \& {Zhu}}{2018}]{bae2018}
{Bae} J.,  {Zhu} Z.,  2018, \mn@doi [\apj] {10.3847/1538-4357/aabf8c}, \href {https://ui.adsabs.harvard.edu/abs/2018ApJ...859..118B} {859, 118}

\bibitem[\protect\citeauthoryear{{Bae}, {Hartmann}  \& {Zhu}}{{Bae} et~al.}{2015}]{bae2015}
{Bae} J.,  {Hartmann} L.,   {Zhu} Z.,  2015, \mn@doi [\apj] {10.1088/0004-637X/805/1/15}, \href {https://ui.adsabs.harvard.edu/abs/2015ApJ...805...15B} {805, 15}

\bibitem[\protect\citeauthoryear{{Bae}, {Isella}, {Zhu}, {Martin}, {Okuzumi}  \& {Suriano}}{{Bae} et~al.}{2023}]{bae2023}
{Bae} J.,  {Isella} A.,  {Zhu} Z.,  {Martin} R.,  {Okuzumi} S.,   {Suriano} S.,  2023, in {Inutsuka} S.,  {Aikawa} Y.,  {Muto} T.,  {Tomida} K.,   {Tamura} M.,  eds,  Astronomical Society of the Pacific Conference Series Vol. 534, Protostars and Planets VII. p.~423 (\mn@eprint {arXiv} {2210.13314}), \mn@doi{10.48550/arXiv.2210.13314}

\bibitem[\protect\citeauthoryear{{Baruteau} et~al.,}{{Baruteau} et~al.}{2019}]{baruteau2019}
{Baruteau} C.,  et~al., 2019, \mn@doi [\mnras] {10.1093/mnras/stz802}, \href {https://ui.adsabs.harvard.edu/abs/2019MNRAS.486..304B} {486, 304}

\bibitem[\protect\citeauthoryear{{Bate}}{{Bate}}{2012}]{bate2012}
{Bate} M.~R.,  2012, \mn@doi [\mnras] {10.1111/j.1365-2966.2011.19955.x}, \href {https://ui.adsabs.harvard.edu/abs/2012MNRAS.419.3115B} {419, 3115}

\bibitem[\protect\citeauthoryear{{Bate}}{{Bate}}{2018}]{bate2018}
{Bate} M.~R.,  2018, \mn@doi [\mnras] {10.1093/mnras/sty169}, \href {https://ui.adsabs.harvard.edu/abs/2018MNRAS.475.5618B} {475, 5618}

\bibitem[\protect\citeauthoryear{{Bate}, {Bonnell}  \& {Price}}{{Bate} et~al.}{1995}]{bate1995}
{Bate} M.~R.,  {Bonnell} I.~A.,   {Price} N.~M.,  1995, \mn@doi [\mnras] {10.1093/mnras/277.2.362}, \href {http://adsabs.harvard.edu/abs/1995MNRAS.277..362B} {277, 362}

\bibitem[\protect\citeauthoryear{{Benisty} et~al.,}{{Benisty} et~al.}{2023}]{benisty2023}
{Benisty} M.,  et~al., 2023, in {Inutsuka} S.,  {Aikawa} Y.,  {Muto} T.,  {Tomida} K.,   {Tamura} M.,  eds,  Astronomical Society of the Pacific Conference Series Vol. 534, Protostars and Planets VII. p.~605 (\mn@eprint {arXiv} {2203.09991}), \mn@doi{10.48550/arXiv.2203.09991}

\bibitem[\protect\citeauthoryear{{Biller} et~al.,}{{Biller} et~al.}{2012}]{biller2012}
{Biller} B.,  et~al., 2012, \mn@doi [\apjl] {10.1088/2041-8205/753/2/L38}, \href {http://adsabs.harvard.edu/abs/2012ApJ...753L..38B} {753, L38}

\bibitem[\protect\citeauthoryear{{Binney} \& {Tremaine}}{{Binney} \& {Tremaine}}{2008}]{binney2008}
{Binney} J.,  {Tremaine} S.,  2008, {Galactic Dynamics: Second Edition}

\bibitem[\protect\citeauthoryear{{Boccaletti} et~al.,}{{Boccaletti} et~al.}{2020}]{boccaletti2020}
{Boccaletti} A.,  et~al., 2020, \mn@doi [\aap] {10.1051/0004-6361/202038008}, \href {https://ui.adsabs.harvard.edu/abs/2020A&A...637L...5B} {637, L5}

\bibitem[\protect\citeauthoryear{{Bondi}}{{Bondi}}{1952}]{bondi1952}
{Bondi} H.,  1952, \mn@doi [\mnras] {10.1093/mnras/112.2.195}, \href {https://ui.adsabs.harvard.edu/abs/1952MNRAS.112..195B} {112, 195}

\bibitem[\protect\citeauthoryear{{Bonnell}, {Smith}, {Clark}  \& {Bate}}{{Bonnell} et~al.}{2011}]{bonnell2011}
{Bonnell} I.~A.,  {Smith} R.~J.,  {Clark} P.~C.,   {Bate} M.~R.,  2011, \mn@doi [\mnras] {10.1111/j.1365-2966.2010.17603.x}, \href {https://ui.adsabs.harvard.edu/abs/2011MNRAS.410.2339B} {410, 2339}

\bibitem[\protect\citeauthoryear{{Borchert}, {Price}, {Pinte}  \& {Cuello}}{{Borchert} et~al.}{2022}]{Borchert2022}
{Borchert} E. M.~A.,  {Price} D.~J.,  {Pinte} C.,   {Cuello} N.,  2022, \mn@doi [\mnras] {10.1093/mnrasl/slab123}, \href {https://ui.adsabs.harvard.edu/abs/2022MNRAS.510L..37B} {510, L37}

\bibitem[\protect\citeauthoryear{{Cacciapuoti} et~al.,}{{Cacciapuoti} et~al.}{2024}]{Cacciapuoti2024}
{Cacciapuoti} L.,  et~al., 2024, \mn@doi [\aap] {10.1051/0004-6361/202347486}, \href {https://ui.adsabs.harvard.edu/abs/2024A&A...682A..61C} {682, A61}

\bibitem[\protect\citeauthoryear{{Calcino}, {Price}, {Pinte}, {van der Marel}, {Ragusa}, {Dipierro}, {Cuello}  \& {Christiaens}}{{Calcino} et~al.}{2019}]{calcino2019}
{Calcino} J.,  {Price} D.~J.,  {Pinte} C.,  {van der Marel} N.,  {Ragusa} E.,  {Dipierro} G.,  {Cuello} N.,   {Christiaens} V.,  2019, \mn@doi [\mnras] {10.1093/mnras/stz2770}, \href {https://ui.adsabs.harvard.edu/abs/2019MNRAS.490.2579C} {490, 2579}

\bibitem[\protect\citeauthoryear{{Calcino}, {Price}, {Pinte}, {Garg}, {Norfolk}, {Christiaens}, {Li}  \& {Teague}}{{Calcino} et~al.}{2023}]{calcino2023}
{Calcino} J.,  {Price} D.~J.,  {Pinte} C.,  {Garg} H.,  {Norfolk} B.~J.,  {Christiaens} V.,  {Li} H.,   {Teague} R.,  2023, \mn@doi [\mnras] {10.1093/mnras/stad1798}, \href {https://ui.adsabs.harvard.edu/abs/2023MNRAS.523.5763C} {523, 5763}

\bibitem[\protect\citeauthoryear{{Calcino} et~al.,}{{Calcino} et~al.}{2024}]{calcino2024}
{Calcino} J.,  et~al., 2024, \mn@doi [arXiv e-prints] {10.48550/arXiv.2407.21309}, \href {https://ui.adsabs.harvard.edu/abs/2024arXiv240721309C} {p. arXiv:2407.21309}

\bibitem[\protect\citeauthoryear{{Cameron}}{{Cameron}}{1978}]{cameron1978}
{Cameron} A.~G.~W.,  1978, \mn@doi [Moon and Planets] {10.1007/BF00896696}, \href {https://ui.adsabs.harvard.edu/abs/1978M&P....18....5C} {18, 5}

\bibitem[\protect\citeauthoryear{{Cassen} \& {Moosman}}{{Cassen} \& {Moosman}}{1981}]{cassen1981}
{Cassen} P.,  {Moosman} A.,  1981, \mn@doi [\icarus] {10.1016/0019-1035(81)90051-8}, \href {https://ui.adsabs.harvard.edu/abs/1981Icar...48..353C} {48, 353}

\bibitem[\protect\citeauthoryear{{Chen}, {Liu}  \& {Han}}{{Chen} et~al.}{2024a}]{chen2024}
{Chen} X.,  {Liu} Z.,   {Han} Z.,  2024a, \mn@doi [Progress in Particle and Nuclear Physics] {10.1016/j.ppnp.2023.104083}, \href {https://ui.adsabs.harvard.edu/abs/2024PrPNP.13404083C} {134, 104083}

\bibitem[\protect\citeauthoryear{{Chen}, {Lawson}, {Brandt}, {Lewis}, {Uyama}, {Millar-Blanchaer}, {Tazaki}  \& {Currie}}{{Chen} et~al.}{2024b}]{chenmh2024}
{Chen} M.,  {Lawson} K.,  {Brandt} T.~D.,  {Lewis} B.~L.,  {Uyama} T.,  {Millar-Blanchaer} M.,  {Tazaki} R.,   {Currie} T.,  2024b, \mn@doi [\mnras] {10.1093/mnras/stae1957}, \href {https://ui.adsabs.harvard.edu/abs/2024MNRAS.533.2473C} {533, 2473}

\bibitem[\protect\citeauthoryear{{Columba} et~al.,}{{Columba} et~al.}{2024}]{columba2024}
{Columba} G.,  et~al., 2024, \mn@doi [\aap] {10.1051/0004-6361/202347109}, \href {https://ui.adsabs.harvard.edu/abs/2024A&A...681A..19C} {681, A19}

\bibitem[\protect\citeauthoryear{{Commer{\c{c}}on}, {Lovascio}, {Lynch}  \& {Ragusa}}{{Commer{\c{c}}on} et~al.}{2024}]{commercon2024}
{Commer{\c{c}}on} B.,  {Lovascio} F.,  {Lynch} E.,   {Ragusa} E.,  2024, \mn@doi [\aap] {10.1051/0004-6361/202449610}, \href {https://ui.adsabs.harvard.edu/abs/2024A&A...689L...9C} {689, L9}

\bibitem[\protect\citeauthoryear{{Cossins}, {Lodato}  \& {Clarke}}{{Cossins} et~al.}{2009}]{cossins2009}
{Cossins} P.,  {Lodato} G.,   {Clarke} C.~J.,  2009, \mn@doi [\mnras] {10.1111/j.1365-2966.2008.14275.x}, \href {https://ui.adsabs.harvard.edu/abs/2009MNRAS.393.1157C} {393, 1157}

\bibitem[\protect\citeauthoryear{{Cuello} et~al.,}{{Cuello} et~al.}{2019}]{cuello2019}
{Cuello} N.,  et~al., 2019, \mn@doi [\mnras] {10.1093/mnras/sty3325}, \href {https://ui.adsabs.harvard.edu/abs/2019MNRAS.483.4114C} {483, 4114}

\bibitem[\protect\citeauthoryear{{Dong}, {Zhu}, {Rafikov}  \& {Stone}}{{Dong} et~al.}{2015}]{dong2015b}
{Dong} R.,  {Zhu} Z.,  {Rafikov} R.~R.,   {Stone} J.~M.,  2015, \mn@doi [\apjl] {10.1088/2041-8205/809/1/L5}, \href {http://adsabs.harvard.edu/abs/2015ApJ...809L...5D} {809, L5}

\bibitem[\protect\citeauthoryear{{Dong} et~al.,}{{Dong} et~al.}{2018}]{dong2018}
{Dong} R.,  et~al., 2018, \mn@doi [\apj] {10.3847/1538-4357/aac6cb}, \href {https://ui.adsabs.harvard.edu/abs/2018ApJ...860..124D} {860, 124}

\bibitem[\protect\citeauthoryear{{Dullemond}, {K{\"u}ffmeier}, {Goicovic}, {Fukagawa}, {Oehl}  \& {Kramer}}{{Dullemond} et~al.}{2019}]{dullemond2019}
{Dullemond} C.~P.,  {K{\"u}ffmeier} M.,  {Goicovic} F.,  {Fukagawa} M.,  {Oehl} V.,   {Kramer} M.,  2019, \mn@doi [\aap] {10.1051/0004-6361/201832632}, \href {https://ui.adsabs.harvard.edu/abs/2019A&A...628A..20D} {628, A20}

\bibitem[\protect\citeauthoryear{{Edgar}}{{Edgar}}{2004}]{edgar2004-review}
{Edgar} R.,  2004, \mn@doi [\nar] {10.1016/j.newar.2004.06.001}, \href {https://ui.adsabs.harvard.edu/abs/2004NewAR..48..843E} {48, 843}

\bibitem[\protect\citeauthoryear{{Elsender}, {Bate}, {Lakeland}, {Jensen}  \& {Lubow}}{{Elsender} et~al.}{2023}]{elsender2023}
{Elsender} D.,  {Bate} M.~R.,  {Lakeland} B.~S.,  {Jensen} E. L.~N.,   {Lubow} S.~H.,  2023, \mn@doi [\mnras] {10.1093/mnras/stad1695}, \href {https://ui.adsabs.harvard.edu/abs/2023MNRAS.523.4353E} {523, 4353}

\bibitem[\protect\citeauthoryear{{Fukagawa} et~al.,}{{Fukagawa} et~al.}{2004}]{fukagawa2004}
{Fukagawa} M.,  et~al., 2004, \mn@doi [\apjl] {10.1086/420699}, \href {https://ui.adsabs.harvard.edu/abs/2004ApJ...605L..53F} {605, L53}

\bibitem[\protect\citeauthoryear{{Fukagawa}, {Tamura}, {Itoh}, {Kudo}, {Imaeda}, {Oasa}, {Hayashi}  \& {Hayashi}}{{Fukagawa} et~al.}{2006}]{fukagawa2006}
{Fukagawa} M.,  {Tamura} M.,  {Itoh} Y.,  {Kudo} T.,  {Imaeda} Y.,  {Oasa} Y.,  {Hayashi} S.~S.,   {Hayashi} M.,  2006, \mn@doi [\apjl] {10.1086/500128}, \href {http://adsabs.harvard.edu/abs/2006ApJ...636L.153F} {636, L153}

\bibitem[\protect\citeauthoryear{{Garg} et~al.,}{{Garg} et~al.}{2021}]{garg2021}
{Garg} H.,  et~al., 2021, \mn@doi [\mnras] {10.1093/mnras/stab800}, \href {https://ui.adsabs.harvard.edu/abs/2021MNRAS.504..782G} {504, 782}

\bibitem[\protect\citeauthoryear{{Garufi} et~al.,}{{Garufi} et~al.}{2020}]{garufi2020}
{Garufi} A.,  et~al., 2020, \mn@doi [\aap] {10.1051/0004-6361/201936946}, \href {https://ui.adsabs.harvard.edu/abs/2020A&A...633A..82G} {633, A82}

\bibitem[\protect\citeauthoryear{{Garufi} et~al.,}{{Garufi} et~al.}{2022}]{garufi2022}
{Garufi} A.,  et~al., 2022, \mn@doi [\aap] {10.1051/0004-6361/202141264}, \href {https://ui.adsabs.harvard.edu/abs/2022A&A...658A.104G} {658, A104}

\bibitem[\protect\citeauthoryear{{Garufi} et~al.,}{{Garufi} et~al.}{2024}]{garufi2024}
{Garufi} A.,  et~al., 2024, \mn@doi [\aap] {10.1051/0004-6361/202347586}, \href {https://ui.adsabs.harvard.edu/abs/2024A&A...685A..53G} {685, A53}

\bibitem[\protect\citeauthoryear{{Ginski} et~al.,}{{Ginski} et~al.}{2021}]{ginksi2021}
{Ginski} C.,  et~al., 2021, \mn@doi [\apjl] {10.3847/2041-8213/abdf57}, \href {https://ui.adsabs.harvard.edu/abs/2021ApJ...908L..25G} {908, L25}

\bibitem[\protect\citeauthoryear{{Grady}, {Woodgate}, {Bruhweiler}, {Boggess}, {Plait}, {Lindler}, {Clampin}  \& {Kalas}}{{Grady} et~al.}{1999}]{grady1999}
{Grady} C.~A.,  {Woodgate} B.,  {Bruhweiler} F.~C.,  {Boggess} A.,  {Plait} P.,  {Lindler} D.~J.,  {Clampin} M.,   {Kalas} P.,  1999, \mn@doi [\apjl] {10.1086/312270}, \href {https://ui.adsabs.harvard.edu/abs/1999ApJ...523L.151G} {523, L151}

\bibitem[\protect\citeauthoryear{{Grady} et~al.,}{{Grady} et~al.}{2001}]{grady2001}
{Grady} C.~A.,  et~al., 2001, \mn@doi [\aj] {10.1086/324447}, \href {https://ui.adsabs.harvard.edu/abs/2001AJ....122.3396G} {122, 3396}

\bibitem[\protect\citeauthoryear{{Gupta} et~al.,}{{Gupta} et~al.}{2023}]{gupta2023}
{Gupta} A.,  et~al., 2023, \mn@doi [\aap] {10.1051/0004-6361/202245254}, \href {https://ui.adsabs.harvard.edu/abs/2023A&A...670L...8G} {670, L8}

\bibitem[\protect\citeauthoryear{{Gupta}, {Miotello}, {Williams}, {Birnstiel}, {Kuffmeier}  \& {Yen}}{{Gupta} et~al.}{2024}]{gupta2024}
{Gupta} A.,  {Miotello} A.,  {Williams} J.~P.,  {Birnstiel} T.,  {Kuffmeier} M.,   {Yen} H.-W.,  2024, \mn@doi [\aap] {10.1051/0004-6361/202348007}, \href {https://ui.adsabs.harvard.edu/abs/2024A&A...683A.133G} {683, A133}

\bibitem[\protect\citeauthoryear{{Haisch}, {Lada}  \& {Lada}}{{Haisch} et~al.}{2001}]{Haisch2001}
{Haisch} Karl~E. J.,  {Lada} E.~A.,   {Lada} C.~J.,  2001, \mn@doi [\apjl] {10.1086/320685}, \href {https://ui.adsabs.harvard.edu/abs/2001ApJ...553L.153H} {553, L153}

\bibitem[\protect\citeauthoryear{{Hales} et~al.,}{{Hales} et~al.}{2024}]{hales2024}
{Hales} A.~S.,  et~al., 2024, \mn@doi [\apj] {10.3847/1538-4357/ad31a1}, \href {https://ui.adsabs.harvard.edu/abs/2024ApJ...966...96H} {966, 96}

\bibitem[\protect\citeauthoryear{{Hall} et~al.,}{{Hall} et~al.}{2020}]{hall2020}
{Hall} C.,  et~al., 2020, arXiv e-prints, \href {https://ui.adsabs.harvard.edu/abs/2020arXiv200715686H} {p. arXiv:2007.15686}

\bibitem[\protect\citeauthoryear{{Hanawa}, {Garufi}, {Podio}, {Codella}  \& {Segura-Cox}}{{Hanawa} et~al.}{2024}]{hanawa2024}
{Hanawa} T.,  {Garufi} A.,  {Podio} L.,  {Codella} C.,   {Segura-Cox} D.,  2024, \mn@doi [\mnras] {10.1093/mnras/stae338}, \href {https://ui.adsabs.harvard.edu/abs/2024MNRAS.528.6581H} {528, 6581}

\bibitem[\protect\citeauthoryear{{Hennebelle}, {Lesur}  \& {Fromang}}{{Hennebelle} et~al.}{2016}]{Hennebelle2016}
{Hennebelle} P.,  {Lesur} G.,   {Fromang} S.,  2016, \mn@doi [\aap] {10.1051/0004-6361/201527877}, \href {https://ui.adsabs.harvard.edu/abs/2016A&A...590A..22H} {590, A22}

\bibitem[\protect\citeauthoryear{{Hennebelle}, {Lesur}  \& {Fromang}}{{Hennebelle} et~al.}{2017}]{Hennebelle2017}
{Hennebelle} P.,  {Lesur} G.,   {Fromang} S.,  2017, \mn@doi [\aap] {10.1051/0004-6361/201629779}, \href {https://ui.adsabs.harvard.edu/abs/2017A&A...599A..86H} {599, A86}

\bibitem[\protect\citeauthoryear{{Hirsh}, {Price}, {Gonzalez}, {Ubeira-Gabellini}  \& {Ragusa}}{{Hirsh} et~al.}{2020}]{hirsh2020}
{Hirsh} K.,  {Price} D.~J.,  {Gonzalez} J.-F.,  {Ubeira-Gabellini} M.~G.,   {Ragusa} E.,  2020, \mn@doi [\mnras] {10.1093/mnras/staa2536}, \href {https://ui.adsabs.harvard.edu/abs/2020MNRAS.498.2936H} {498, 2936}

\bibitem[\protect\citeauthoryear{{Huang} et~al.,}{{Huang} et~al.}{2020}]{huang2020}
{Huang} J.,  et~al., 2020, \mn@doi [\apj] {10.3847/1538-4357/aba1e1}, \href {https://ui.adsabs.harvard.edu/abs/2020ApJ...898..140H} {898, 140}

\bibitem[\protect\citeauthoryear{{Huang} et~al.,}{{Huang} et~al.}{2021}]{huang2021}
{Huang} J.,  et~al., 2021, \mn@doi [\apjs] {10.3847/1538-4365/ac143e}, \href {https://ui.adsabs.harvard.edu/abs/2021ApJS..257...19H} {257, 19}

\bibitem[\protect\citeauthoryear{{Huang} et~al.,}{{Huang} et~al.}{2022}]{huang2022}
{Huang} J.,  et~al., 2022, \mn@doi [\apj] {10.3847/1538-4357/ac63ba}, \href {https://ui.adsabs.harvard.edu/abs/2022ApJ...930..171H} {930, 171}

\bibitem[\protect\citeauthoryear{{Hunziker} et~al.,}{{Hunziker} et~al.}{2021}]{Hunziker2021}
{Hunziker} S.,  et~al., 2021, \mn@doi [\aap] {10.1051/0004-6361/202040166}, \href {https://ui.adsabs.harvard.edu/abs/2021A&A...648A.110H} {648, A110}

\bibitem[\protect\citeauthoryear{{Juillard}, {Christiaens}  \& {Absil}}{{Juillard} et~al.}{2022}]{Juillard2022}
{Juillard} S.,  {Christiaens} V.,   {Absil} O.,  2022, \mn@doi [\aap] {10.1051/0004-6361/202244402}, \href {https://ui.adsabs.harvard.edu/abs/2022A&A...668A.125J} {668, A125}

\bibitem[\protect\citeauthoryear{{Klessen} \& {Hennebelle}}{{Klessen} \& {Hennebelle}}{2010}]{klessen2010}
{Klessen} R.~S.,  {Hennebelle} P.,  2010, \mn@doi [\aap] {10.1051/0004-6361/200913780}, \href {https://ui.adsabs.harvard.edu/abs/2010A&A...520A..17K} {520, A17}

\bibitem[\protect\citeauthoryear{{Krieger}, {Kuffmeier}, {Reissl}, {Dullemond}, {Ginski}  \& {Wolf}}{{Krieger} et~al.}{2024}]{Krieger2024}
{Krieger} A.,  {Kuffmeier} M.,  {Reissl} S.,  {Dullemond} C.~P.,  {Ginski} C.,   {Wolf} S.,  2024, \mn@doi [\aap] {10.1051/0004-6361/202348354}, \href {https://ui.adsabs.harvard.edu/abs/2024A&A...686A.111K} {686, A111}

\bibitem[\protect\citeauthoryear{{Kuffmeier}}{{Kuffmeier}}{2024}]{kuffmeier2024}
{Kuffmeier} M.,  2024, \mn@doi [Frontiers in Astronomy and Space Sciences] {10.3389/fspas.2024.1403075}, \href {https://ui.adsabs.harvard.edu/abs/2024FrASS..1103075K} {11, 1403075}

\bibitem[\protect\citeauthoryear{{Kuffmeier}, {Haugb{\o}lle}  \& {Nordlund}}{{Kuffmeier} et~al.}{2017}]{kuffmeier2017}
{Kuffmeier} M.,  {Haugb{\o}lle} T.,   {Nordlund} {\r{A}}.,  2017, \mn@doi [\apj] {10.3847/1538-4357/aa7c64}, \href {https://ui.adsabs.harvard.edu/abs/2017ApJ...846....7K} {846, 7}

\bibitem[\protect\citeauthoryear{{Kuffmeier}, {Frimann}, {Jensen}  \& {Haugb{\o}lle}}{{Kuffmeier} et~al.}{2018}]{kuffmeier2018}
{Kuffmeier} M.,  {Frimann} S.,  {Jensen} S.~S.,   {Haugb{\o}lle} T.,  2018, \mn@doi [\mnras] {10.1093/mnras/sty024}, \href {https://ui.adsabs.harvard.edu/abs/2018MNRAS.475.2642K} {475, 2642}

\bibitem[\protect\citeauthoryear{{Kuffmeier}, {Goicovic}  \& {Dullemond}}{{Kuffmeier} et~al.}{2020}]{Kuffmeier2020}
{Kuffmeier} M.,  {Goicovic} F.~G.,   {Dullemond} C.~P.,  2020, \mn@doi [\aap] {10.1051/0004-6361/201936820}, \href {https://ui.adsabs.harvard.edu/abs/2020A&A...633A...3K} {633, A3}

\bibitem[\protect\citeauthoryear{{Kuffmeier}, {Dullemond}, {Reissl}  \& {Goicovic}}{{Kuffmeier} et~al.}{2021}]{kuffmeier2021}
{Kuffmeier} M.,  {Dullemond} C.~P.,  {Reissl} S.,   {Goicovic} F.~G.,  2021, \mn@doi [\aap] {10.1051/0004-6361/202039614}, \href {https://ui.adsabs.harvard.edu/abs/2021A&A...656A.161K} {656, A161}

\bibitem[\protect\citeauthoryear{{Kuffmeier}, {Jensen}  \& {Haugb{\o}lle}}{{Kuffmeier} et~al.}{2023}]{kuffmeier2023}
{Kuffmeier} M.,  {Jensen} S.~S.,   {Haugb{\o}lle} T.,  2023, \mn@doi [European Physical Journal Plus] {10.1140/epjp/s13360-023-03880-y}, \href {https://ui.adsabs.harvard.edu/abs/2023EPJP..138..272K} {138, 272}

\bibitem[\protect\citeauthoryear{{Kuffmeier}, {Pineda}, {Segura-Cox}  \& {Haugb{\o}lle}}{{Kuffmeier} et~al.}{2024}]{kuffmeier2024b}
{Kuffmeier} M.,  {Pineda} J.~E.,  {Segura-Cox} D.,   {Haugb{\o}lle} T.,  2024, \mn@doi [\aap] {10.1051/0004-6361/202450410}, \href {https://ui.adsabs.harvard.edu/abs/2024A&A...690A.297K} {690, A297}

\bibitem[\protect\citeauthoryear{{Kuo}, {Yen}, {Gu}  \& {Chang}}{{Kuo} et~al.}{2022}]{kuo2022}
{Kuo} I. H.~G.,  {Yen} H.-W.,  {Gu} P.-G.,   {Chang} T.-E.,  2022, \mn@doi [\apj] {10.3847/1538-4357/ac9228}, \href {https://ui.adsabs.harvard.edu/abs/2022ApJ...938...50K} {938, 50}

\bibitem[\protect\citeauthoryear{{Kuznetsova}, {Bae}, {Hartmann}  \& {Mac Low}}{{Kuznetsova} et~al.}{2022}]{Kuznetsova2022}
{Kuznetsova} A.,  {Bae} J.,  {Hartmann} L.,   {Mac Low} M.-M.,  2022, \mn@doi [\apj] {10.3847/1538-4357/ac54a8}, \href {https://ui.adsabs.harvard.edu/abs/2022ApJ...928...92K} {928, 92}

\bibitem[\protect\citeauthoryear{{Law} et~al.,}{{Law} et~al.}{2021}]{law2021}
{Law} C.~J.,  et~al., 2021, arXiv e-prints, \href {https://ui.adsabs.harvard.edu/abs/2021arXiv210906217L} {p. arXiv:2109.06217}

\bibitem[\protect\citeauthoryear{{Lesur}, {Hennebelle}  \& {Fromang}}{{Lesur} et~al.}{2015}]{lesur2015}
{Lesur} G.,  {Hennebelle} P.,   {Fromang} S.,  2015, \mn@doi [\aap] {10.1051/0004-6361/201526734}, \href {https://ui.adsabs.harvard.edu/abs/2015A&A...582L...9L} {582, L9}

\bibitem[\protect\citeauthoryear{{Lin} \& {Pringle}}{{Lin} \& {Pringle}}{1990}]{lin1990}
{Lin} D. N.~C.,  {Pringle} J.~E.,  1990, \mn@doi [\apj] {10.1086/169004}, \href {https://ui.adsabs.harvard.edu/abs/1990ApJ...358..515L} {358, 515}

\bibitem[\protect\citeauthoryear{{Liu} et~al.,}{{Liu} et~al.}{2016}]{liu2016}
{Liu} H.~B.,  et~al., 2016, \mn@doi [Science Advances] {10.1126/sciadv.1500875}, \href {https://ui.adsabs.harvard.edu/abs/2016SciA....2E0875L} {2, e1500875}

\bibitem[\protect\citeauthoryear{{Lodato} \& {Price}}{{Lodato} \& {Price}}{2010}]{lodato2010}
{Lodato} G.,  {Price} D.~J.,  2010, \mn@doi [\mnras] {10.1111/j.1365-2966.2010.16526.x}, \href {https://ui.adsabs.harvard.edu/abs/2010MNRAS.405.1212L} {405, 1212}

\bibitem[\protect\citeauthoryear{{Longarini}, {Lodato}, {Toci}, {Veronesi}, {Hall}, {Dong}  \& {Patrick Terry}}{{Longarini} et~al.}{2021}]{Longarini2021}
{Longarini} C.,  {Lodato} G.,  {Toci} C.,  {Veronesi} B.,  {Hall} C.,  {Dong} R.,   {Patrick Terry} J.,  2021, \mn@doi [\apjl] {10.3847/2041-8213/ac2df6}, \href {https://ui.adsabs.harvard.edu/abs/2021ApJ...920L..41L} {920, L41}

\bibitem[\protect\citeauthoryear{{Mamajek}}{{Mamajek}}{2009}]{mamajek2009}
{Mamajek} E.~E.,  2009, in {Usuda} T.,  {Tamura} M.,   {Ishii} M.,  eds,  American Institute of Physics Conference Series Vol. 1158, Exoplanets and Disks: Their Formation and Diversity. AIP, pp 3--10 (\mn@eprint {arXiv} {0906.5011}), \mn@doi{10.1063/1.3215910}

\bibitem[\protect\citeauthoryear{{McCrea}}{{McCrea}}{1960}]{mccrea1960}
{McCrea} W.~H.,  1960, \mn@doi [Proceedings of the Royal Society of London Series A] {10.1098/rspa.1960.0108}, \href {https://ui.adsabs.harvard.edu/abs/1960RSPSA.256..245M} {256, 245}

\bibitem[\protect\citeauthoryear{{Mesa} et~al.,}{{Mesa} et~al.}{2022}]{Mesa2022}
{Mesa} D.,  et~al., 2022, \mn@doi [\aap] {10.1051/0004-6361/202142219}, \href {https://ui.adsabs.harvard.edu/abs/2022A&A...658A..63M} {658, A63}

\bibitem[\protect\citeauthoryear{{Monaghan}}{{Monaghan}}{1992}]{monaghan1992}
{Monaghan} J.~J.,  1992, \mn@doi [\araa] {10.1146/annurev.aa.30.090192.002551}, \href {https://ui.adsabs.harvard.edu/abs/1992ARA&A..30..543M} {30, 543}

\bibitem[\protect\citeauthoryear{{Monnier} et~al.,}{{Monnier} et~al.}{2019}]{monnier2019}
{Monnier} J.~D.,  et~al., 2019, \mn@doi [\apj] {10.3847/1538-4357/aafe87}, \href {https://ui.adsabs.harvard.edu/abs/2019ApJ...872..122M} {872, 122}

\bibitem[\protect\citeauthoryear{{Norfolk} et~al.,}{{Norfolk} et~al.}{2022}]{norfolk2022}
{Norfolk} B.~J.,  et~al., 2022, \mn@doi [\apjl] {10.3847/2041-8213/ac85ed}, \href {https://ui.adsabs.harvard.edu/abs/2022ApJ...936L...4N} {936, L4}

\bibitem[\protect\citeauthoryear{{Nowak}, {Rowther}, {Lacour}, {Meru}, {Nealon}  \& {Price}}{{Nowak} et~al.}{2024}]{nowak2024}
{Nowak} M.,  {Rowther} S.,  {Lacour} S.,  {Meru} F.,  {Nealon} R.,   {Price} D.~J.,  2024, \mn@doi [\aap] {10.1051/0004-6361/202347748}, \href {https://ui.adsabs.harvard.edu/abs/2024A&A...683A...6N} {683, A6}

\bibitem[\protect\citeauthoryear{{Ogilvie} \& {Barker}}{{Ogilvie} \& {Barker}}{2014}]{ogilvie2014}
{Ogilvie} G.~I.,  {Barker} A.~J.,  2014, \mn@doi [\mnras] {10.1093/mnras/stu1795}, \href {https://ui.adsabs.harvard.edu/abs/2014MNRAS.445.2621O} {445, 2621}

\bibitem[\protect\citeauthoryear{{Padoan}, {Kritsuk}, {Norman}  \& {Nordlund}}{{Padoan} et~al.}{2005}]{padoan2005}
{Padoan} P.,  {Kritsuk} A.,  {Norman} M.~L.,   {Nordlund} {\r{A}}.,  2005, \mn@doi [\apjl] {10.1086/429562}, \href {https://ui.adsabs.harvard.edu/abs/2005ApJ...622L..61P} {622, L61}

\bibitem[\protect\citeauthoryear{{Padoan}, {Pan}, {Juvela}, {Haugb{\o}lle}  \& {Nordlund}}{{Padoan} et~al.}{2020}]{padoan2020}
{Padoan} P.,  {Pan} L.,  {Juvela} M.,  {Haugb{\o}lle} T.,   {Nordlund} {\r{A}}.,  2020, \mn@doi [\apj] {10.3847/1538-4357/abaa47}, \href {https://ui.adsabs.harvard.edu/abs/2020ApJ...900...82P} {900, 82}

\bibitem[\protect\citeauthoryear{{Padoan}, {Pan}, {Pelkonen}, {Haugboelle}  \& {Nordlund}}{{Padoan} et~al.}{2024}]{Padoan2024}
{Padoan} P.,  {Pan} L.,  {Pelkonen} V.-M.,  {Haugboelle} T.,   {Nordlund} A.,  2024, \mn@doi [arXiv e-prints] {10.48550/arXiv.2405.07334}, \href {https://ui.adsabs.harvard.edu/abs/2024arXiv240507334P} {p. arXiv:2405.07334}

\bibitem[\protect\citeauthoryear{{Pelkonen}, {Padoan}, {Haugb{\o}lle}  \& {Nordlund}}{{Pelkonen} et~al.}{2021}]{pelkonen2021}
{Pelkonen} V.~M.,  {Padoan} P.,  {Haugb{\o}lle} T.,   {Nordlund} {\r{A}}.,  2021, \mn@doi [\mnras] {10.1093/mnras/stab844}, \href {https://ui.adsabs.harvard.edu/abs/2021MNRAS.504.1219P} {504, 1219}

\bibitem[\protect\citeauthoryear{{Pelkonen}, {Padoan}, {Juvela}, {Haugb{\o}lle}  \& {Nordlund}}{{Pelkonen} et~al.}{2024}]{pelkonen2024}
{Pelkonen} V.-M.,  {Padoan} P.,  {Juvela} M.,  {Haugb{\o}lle} T.,   {Nordlund} {\r{A}}.,  2024, \mn@doi [arXiv e-prints] {10.48550/arXiv.2405.06520}, \href {https://ui.adsabs.harvard.edu/abs/2024arXiv240506520P} {p. arXiv:2405.06520}

\bibitem[\protect\citeauthoryear{{Pineda}, {Segura-Cox}, {Caselli}, {Cunningham}, {Zhao}, {Schmiedeke}, {Maureira}  \& {Neri}}{{Pineda} et~al.}{2020}]{pineda2020}
{Pineda} J.~E.,  {Segura-Cox} D.,  {Caselli} P.,  {Cunningham} N.,  {Zhao} B.,  {Schmiedeke} A.,  {Maureira} M.~J.,   {Neri} R.,  2020, \mn@doi [Nature Astronomy] {10.1038/s41550-020-1150-z}, \href {https://ui.adsabs.harvard.edu/abs/2020NatAs...4.1158P} {4, 1158}

\bibitem[\protect\citeauthoryear{{Pinte}, {Teague}, {Flaherty}, {Hall}, {Facchini}  \& {Casassus}}{{Pinte} et~al.}{2023}]{pinte2023}
{Pinte} C.,  {Teague} R.,  {Flaherty} K.,  {Hall} C.,  {Facchini} S.,   {Casassus} S.,  2023, in {Inutsuka} S.,  {Aikawa} Y.,  {Muto} T.,  {Tomida} K.,   {Tamura} M.,  eds,  Astronomical Society of the Pacific Conference Series Vol. 534, Protostars and Planets VII. p.~645 (\mn@eprint {arXiv} {2203.09528}), \mn@doi{10.48550/arXiv.2203.09528}

\bibitem[\protect\citeauthoryear{{Poblete}, {Calcino}, {Cuello}, {Mac{\'\i}as}, {Ribas}, {Price}, {Cuadra}  \& {Pinte}}{{Poblete} et~al.}{2020}]{poblete2020}
{Poblete} P.~P.,  {Calcino} J.,  {Cuello} N.,  {Mac{\'\i}as} E.,  {Ribas} {\'A}.,  {Price} D.~J.,  {Cuadra} J.,   {Pinte} C.,  2020, \mn@doi [\mnras] {10.1093/mnras/staa1655}, \href {https://ui.adsabs.harvard.edu/abs/2020MNRAS.tmp.1807P} {}

\bibitem[\protect\citeauthoryear{{Price} et~al.,}{{Price} et~al.}{2018a}]{phantom2018}
{Price} D.~J.,  et~al., 2018a, \mn@doi [\pasa] {10.1017/pasa.2018.25}, \href {https://ui.adsabs.harvard.edu/abs/2018PASA...35...31P} {35, e031}

\bibitem[\protect\citeauthoryear{{Price} et~al.,}{{Price} et~al.}{2018b}]{price2018}
{Price} D.~J.,  et~al., 2018b, \mn@doi [\mnras] {10.1093/mnras/sty647}, \href {http://adsabs.harvard.edu/abs/2018MNRAS.477.1270P} {477, 1270}

\bibitem[\protect\citeauthoryear{{Pringle}}{{Pringle}}{1981}]{pringle1981}
{Pringle} J.~E.,  1981, \mn@doi [\araa] {10.1146/annurev.aa.19.090181.001033}, \href {https://ui.adsabs.harvard.edu/abs/1981ARA&A..19..137P} {19, 137}

\bibitem[\protect\citeauthoryear{{Ragusa}, {Dipierro}, {Lodato}, {Laibe}  \& {Price}}{{Ragusa} et~al.}{2017a}]{ragusa2017}
{Ragusa} E.,  {Dipierro} G.,  {Lodato} G.,  {Laibe} G.,   {Price} D.~J.,  2017a, \mn@doi [\mnras] {10.1093/mnras/stw2456}, \href {http://adsabs.harvard.edu/abs/2017MNRAS.464.1449R} {464, 1449}

\bibitem[\protect\citeauthoryear{Ragusa, Rosotti, Teyssandier, Booth, Clarke  \& Lodato}{Ragusa et~al.}{2017b}]{ragusa2018}
Ragusa E.,  Rosotti G.,  Teyssandier J.,  Booth R.,  Clarke C.~J.,   Lodato G.,  2017b, \mn@doi [Monthly Notices of the Royal Astronomical Society] {10.1093/mnras/stx3094}, 474, 4460

\bibitem[\protect\citeauthoryear{{Ragusa}, {Lynch}, {Laibe}, {Longarini}  \& {Ceppi}}{{Ragusa} et~al.}{2024}]{ragusa2024}
{Ragusa} E.,  {Lynch} E.,  {Laibe} G.,  {Longarini} C.,   {Ceppi} S.,  2024, \mn@doi [\aap] {10.1051/0004-6361/202449583}, \href {https://ui.adsabs.harvard.edu/abs/2024A&A...686A.264R} {686, A264}

\bibitem[\protect\citeauthoryear{{Reggiani} et~al.,}{{Reggiani} et~al.}{2018}]{reggiani2018}
{Reggiani} M.,  et~al., 2018, \mn@doi [\aap] {10.1051/0004-6361/201732016}, \href {http://adsabs.harvard.edu/abs/2018A%26A...611A..74R} {611, A74}

\bibitem[\protect\citeauthoryear{{Ren} et~al.,}{{Ren} et~al.}{2018}]{ren2018}
{Ren} B.,  et~al., 2018, \mn@doi [\apj] {10.3847/2041-8213/aab7f5}, \href {https://ui.adsabs.harvard.edu/#abs/2018ApJ...857L...9R} {857, L9}

\bibitem[\protect\citeauthoryear{{Ren} et~al.,}{{Ren} et~al.}{2020}]{ren2020}
{Ren} B.,  et~al., 2020, arXiv e-prints, \href {https://ui.adsabs.harvard.edu/abs/2020arXiv200704980R} {p. arXiv:2007.04980}

\bibitem[\protect\citeauthoryear{{Ren} et~al.,}{{Ren} et~al.}{2024}]{ren2024}
{Ren} B.~B.,  et~al., 2024, \mn@doi [\aap] {10.1051/0004-6361/202348114}, \href {https://ui.adsabs.harvard.edu/abs/2024A&A...681L...2R} {681, L2}

\bibitem[\protect\citeauthoryear{{Shakura} \& {Sunyaev}}{{Shakura} \& {Sunyaev}}{1973}]{shakura1973}
{Shakura} N.~I.,  {Sunyaev} R.~A.,  1973, \aap, \href {http://adsabs.harvard.edu/abs/1973A%26A....24..337S} {24, 337}

\bibitem[\protect\citeauthoryear{{Smallwood}, {Yang}, {Zhu}, {Martin}, {Dong}, {Cuello}  \& {Isella}}{{Smallwood} et~al.}{2023}]{smallwood2023}
{Smallwood} J.~L.,  {Yang} C.-C.,  {Zhu} Z.,  {Martin} R.~G.,  {Dong} R.,  {Cuello} N.,   {Isella} A.,  2023, \mn@doi [\mnras] {10.1093/mnras/stad742}, \href {https://ui.adsabs.harvard.edu/abs/2023MNRAS.521.3500S} {521, 3500}

\bibitem[\protect\citeauthoryear{{Smith}, {Glover}, {Bonnell}, {Clark}  \& {Klessen}}{{Smith} et~al.}{2011}]{smith2011}
{Smith} R.~J.,  {Glover} S. C.~O.,  {Bonnell} I.~A.,  {Clark} P.~C.,   {Klessen} R.~S.,  2011, \mn@doi [\mnras] {10.1111/j.1365-2966.2010.17775.x}, \href {https://ui.adsabs.harvard.edu/abs/2011MNRAS.411.1354S} {411, 1354}

\bibitem[\protect\citeauthoryear{{Speedie} et~al.,}{{Speedie} et~al.}{2024}]{speedie2024}
{Speedie} J.,  et~al., 2024, \mn@doi [arXiv e-prints] {10.48550/arXiv.2409.02196}, \href {https://ui.adsabs.harvard.edu/abs/2024arXiv240902196S} {p. arXiv:2409.02196}

\bibitem[\protect\citeauthoryear{{Stadler} et~al.,}{{Stadler} et~al.}{2023}]{stadler2023}
{Stadler} J.,  et~al., 2023, \mn@doi [\aap] {10.1051/0004-6361/202245381}, \href {https://ui.adsabs.harvard.edu/abs/2023A&A...670L...1S} {670, L1}

\bibitem[\protect\citeauthoryear{{Teyssandier} \& {Ogilvie}}{{Teyssandier} \& {Ogilvie}}{2016}]{Teyssandier2016}
{Teyssandier} J.,  {Ogilvie} G.~I.,  2016, \mn@doi [\mnras] {10.1093/mnras/stw521}, \href {https://ui.adsabs.harvard.edu/abs/2016MNRAS.458.3221T} {458, 3221}

\bibitem[\protect\citeauthoryear{{Teyssandier} \& {Ogilvie}}{{Teyssandier} \& {Ogilvie}}{2017}]{Teyssandier2017}
{Teyssandier} J.,  {Ogilvie} G.~I.,  2017, \mn@doi [\mnras] {10.1093/mnras/stx426}, \href {https://ui.adsabs.harvard.edu/abs/2017MNRAS.467.4577T} {467, 4577}

\bibitem[\protect\citeauthoryear{{Thies}, {Kroupa}, {Goodwin}, {Stamatellos}  \& {Whitworth}}{{Thies} et~al.}{2011}]{thies2011}
{Thies} I.,  {Kroupa} P.,  {Goodwin} S.~P.,  {Stamatellos} D.,   {Whitworth} A.~P.,  2011, \mn@doi [\mnras] {10.1111/j.1365-2966.2011.19390.x}, \href {https://ui.adsabs.harvard.edu/abs/2011MNRAS.417.1817T} {417, 1817}

\bibitem[\protect\citeauthoryear{{Throop} \& {Bally}}{{Throop} \& {Bally}}{2008}]{throop2008}
{Throop} H.~B.,  {Bally} J.,  2008, \mn@doi [\aj] {10.1088/0004-6256/135/6/2380}, \href {https://ui.adsabs.harvard.edu/abs/2008AJ....135.2380T} {135, 2380}

\bibitem[\protect\citeauthoryear{{Ulrich}}{{Ulrich}}{1976}]{ulrich1976}
{Ulrich} R.~K.,  1976, \mn@doi [\apj] {10.1086/154840}, \href {https://ui.adsabs.harvard.edu/abs/1976ApJ...210..377U} {210, 377}

\bibitem[\protect\citeauthoryear{{Unno}, {Hanawa}  \& {Takasao}}{{Unno} et~al.}{2022}]{unno2022}
{Unno} M.,  {Hanawa} T.,   {Takasao} S.,  2022, \mn@doi [\apj] {10.3847/1538-4357/aca410}, \href {https://ui.adsabs.harvard.edu/abs/2022ApJ...941..154U} {941, 154}

\bibitem[\protect\citeauthoryear{{Uyama} et~al.,}{{Uyama} et~al.}{2020}]{Uyama2020b}
{Uyama} T.,  et~al., 2020, \mn@doi [\apj] {10.3847/1538-4357/aba8f6}, \href {https://ui.adsabs.harvard.edu/abs/2020ApJ...900..135U} {900, 135}

\bibitem[\protect\citeauthoryear{{Valdivia-Mena} et~al.,}{{Valdivia-Mena} et~al.}{2024}]{Valdivia-Mena2024}
{Valdivia-Mena} M.~T.,  et~al., 2024, \mn@doi [\aap] {10.1051/0004-6361/202449395}, \href {https://ui.adsabs.harvard.edu/abs/2024A&A...687A..71V} {687, A71}

\bibitem[\protect\citeauthoryear{{Wagner}, {Stone}, {Spalding}, {Apai}, {Dong}, {Ertel}, {Leisenring}  \& {Webster}}{{Wagner} et~al.}{2019}]{wagner2019}
{Wagner} K.,  {Stone} J.~M.,  {Spalding} E.,  {Apai} D.,  {Dong} R.,  {Ertel} S.,  {Leisenring} J.,   {Webster} R.,  2019, \mn@doi [\apj] {10.3847/1538-4357/ab32ea}, \href {https://ui.adsabs.harvard.edu/abs/2019ApJ...882...20W} {882, 20}

\bibitem[\protect\citeauthoryear{{Wagner} et~al.,}{{Wagner} et~al.}{2023}]{wagner2023}
{Wagner} K.,  et~al., 2023, \mn@doi [Nature Astronomy] {10.1038/s41550-023-02028-3}, \href {https://ui.adsabs.harvard.edu/abs/2023NatAs...7.1208W} {7, 1208}

\bibitem[\protect\citeauthoryear{{Wijnen}, {Pelupessy}, {Pols}  \& {Portegies Zwart}}{{Wijnen} et~al.}{2017}]{wijnen2017}
{Wijnen} T.~P.~G.,  {Pelupessy} F.~I.,  {Pols} O.~R.,   {Portegies Zwart} S.,  2017, \mn@doi [\aap] {10.1051/0004-6361/201730793}, \href {https://ui.adsabs.harvard.edu/abs/2017A&A...604A..88W} {604, A88}

\bibitem[\protect\citeauthoryear{{Winter}, {Benisty}  \& {Andrews}}{{Winter} et~al.}{2024}]{winter2024}
{Winter} A.~J.,  {Benisty} M.,   {Andrews} S.~M.,  2024, \mn@doi [\apjl] {10.3847/2041-8213/ad6d5d}, \href {https://ui.adsabs.harvard.edu/abs/2024ApJ...972L...9W} {972, L9}

\bibitem[\protect\citeauthoryear{{Xie}, {Ren}, {Dong}, {Pueyo}, {Ruffio}, {Fang}, {Mawet}  \& {Stolker}}{{Xie} et~al.}{2021}]{xie2021}
{Xie} C.,  {Ren} B.,  {Dong} R.,  {Pueyo} L.,  {Ruffio} J.-B.,  {Fang} T.,  {Mawet} D.,   {Stolker} T.,  2021, \mn@doi [\apjl] {10.3847/2041-8213/abd241}, \href {https://ui.adsabs.harvard.edu/abs/2021ApJ...906L...9X} {906, L9}

\bibitem[\protect\citeauthoryear{{Xie} et~al.,}{{Xie} et~al.}{2023}]{xie2023}
{Xie} C.,  et~al., 2023, \mn@doi [\aap] {10.1051/0004-6361/202346305}, \href {https://ui.adsabs.harvard.edu/abs/2023A&A...675L...1X} {675, L1}

\bibitem[\protect\citeauthoryear{{Yang}, {Fern{\'a}ndez-L{\'o}pez}, {Li}, {Stephens}, {Looney}, {Lin}  \& {Harrison}}{{Yang} et~al.}{2023}]{yang2023}
{Yang} H.,  {Fern{\'a}ndez-L{\'o}pez} M.,  {Li} Z.-Y.,  {Stephens} I.~W.,  {Looney} L.~W.,  {Lin} Z.-Y.~D.,   {Harrison} R.,  2023, \mn@doi [\apjl] {10.3847/2041-8213/acccf8}, \href {https://ui.adsabs.harvard.edu/abs/2023ApJ...948L...2Y} {948, L2}

\bibitem[\protect\citeauthoryear{{Zhu}, {Stone}, {Rafikov}  \& {Bai}}{{Zhu} et~al.}{2014}]{zhu2014}
{Zhu} Z.,  {Stone} J.~M.,  {Rafikov} R.~R.,   {Bai} X.-n.,  2014, \mn@doi [\apj] {10.1088/0004-637X/785/2/122}, \href {http://adsabs.harvard.edu/abs/2014ApJ...785..122Z} {785, 122}

\bibitem[\protect\citeauthoryear{{van der Marel}, {van Dishoeck}, {Bruderer}, {P{\'e}rez}  \& {Isella}}{{van der Marel} et~al.}{2015}]{vandermarel2015b}
{van der Marel} N.,  {van Dishoeck} E.~F.,  {Bruderer} S.,  {P{\'e}rez} L.,   {Isella} A.,  2015, \mn@doi [\aap] {10.1051/0004-6361/201525658}, \href {https://ui.adsabs.harvard.edu/abs/2015A&A...579A.106V} {579, A106}

\makeatother
\end{thebibliography}

%%%%%%%%%%%%%%%%%%%%%%%%%%%%%%%%%%%%%%%%%%%%%%%%%%

%%%%%%%%%%%%%%%%% APPENDICES %%%%%%%%%%%%%%%%%%%%%

% \appendix

%%%%%%%%%%%%%%%%%%%%%%%%%%%%%%%%%%%%%%%%%%%%%%%%%%

% Don't change these lines
\bsp	% typesetting comment
\label{lastpage}
\end{document}